\documentclass[trackchanges,twocolumn, twocolappendix]{aastex7}
\usepackage{graphicx} 

\usepackage[version=3]{mhchem}
\usepackage{amsmath,amstext}
\usepackage{amssymb}
\usepackage{mysymbol}
\usepackage{comment}
\usepackage{xcolor} 

\newcommand{\bl}[1]{\mbox{\boldmath$ #1 $}}

\date{Accepted Oct 25, 2025}

\begin{document}

\title{Morphological and Kinematical Diagnostic of FU Orionis-type Outburst Mechanisms}

\author[0000-0003-4361-5577]{Jinshi Sai}
\affiliation{Academia Sinica Institute of Astronomy \& Astrophysics (ASIAA), 11F of Astronomy-Mathematics Building, AS/NTU, No.1, Sec. 4, Roosevelt Rd, Taipei 106319, Taiwan}
\affiliation{Department of Physics and Astronomy, Graduate School of Science and Engineering, Kagoshima University, 1-21-35 Korimoto, Kagoshima, Kagoshima 890-0065, Japan}
\email{jn.insa.sai@gmail.com}
\email[show]{jinshi.sai@sci.kagoshima-u.ac.jp}

\author[0000-0002-6045-0359]{Eduard I. Vorobyov}
\affiliation{University of Vienna, Department of Astrophysics, T\"urkenschanzstrasse 17, 1180, Vienna, Austria}
\affiliation{Research Institute of Physics, Southern Federal University, Rostov-on-Don 344090, Russia}
\email{eduard.vorobiev@univie.ac.at}

\author[0000-0002-2479-3370]{Alexandr Skliarevskii}
\affiliation{Research Institute of Physics, Southern Federal University, Rostov-on-Don 344090, Russia}
\email{sklyarevskiy@sfedu.ru}

\author[0000-0001-9248-7546]{Michihiro Takami}
\affiliation{Academia Sinica Institute of Astronomy \& Astrophysics (ASIAA), 11F of Astronomy-Mathematics Building, AS/NTU, No.1, Sec. 4, Roosevelt Rd, Taipei 106319, Taiwan}
\email{hiro@asiaa.sinica.edu.tw}

\correspondingauthor{Jinshi Sai}

\keywords{\uat{Star formation}{1569} --- \uat{FU Orionis stars}{553} --- \uat{Young stellar objects}{1834} --- \uat{Protostars}{1302} --- \uat{Circumstellar disks}{235} --- \uat{Protoplanetary disks}{1300} --- \uat{Stellar accretion disks}{1579}}

\begin{abstract}
We investigated the possibility of determining the mechanism of the FU Orionis-type outburst based on molecular line observations of protoplanetary disks with synthetic observations of distinct numerical burst models. The morphology of the synthetic \ce{C^18O} emission is sensitive to gas temperature and does not coincide with the actual gas disk structures, particularly in the magnetorotational instability (MRI) and clump-infall models, which exhibit peculiar temperature distributions. This highlights the need for careful interpretation of morphologies of line emission from disks under accretion outbursts. The synthetic \ce{C^18O} emission of each model exhibits distinct kinematic features that can be used to distinguish outburst scenarios. In the MRI model, kinematic features of the gravitational instability (GI), which fuels MRI-driven accretion bursts, are small in both amplitude and spatial extent, resulting in no prominent local features in the residual velocity map at a typical distance for FU Orionis-type objects. In contrast, the clump-infall model shows a clear sign of gas expansion along a spiral, which is caused by exchange of angular momentum between an infalling clump and surrounding gas. The intruder model exhibits a highly asymmetric velocity structure with respect to the systemic velocity of the primary protostar in velocity channel maps. These distinct kinematic features may serve as promising diagnostics for distinguishing the physical mechanisms responsible for FU Orionis-type outbursts.
\end{abstract}

 \section{Introduction}

Stellar mass is the most fundamental quantity for stars, characterizing their various properties and lives. 
However, the process by which low-mass stars acquire their mass remains an open question. Early observations found that the luminosity of protostars tends to be significantly lower than theoretical predictions for steady mass accretion by orders of magnitudes, which is the so-called luminosity problem \citep[][see also \citealt{Audard:2014aa, Fischer:2023aa} for reviews]{Kenyon1990}. One of the solutions for this long-standing issue is episodic accretion \citep{Dunham2012}, where mass accretion onto a protostar is time variable. In this scenario, protostars experience rapid growth through short-term accretion outbursts, while spending most of their lifetimes in quiescent phases.

FU Orionis type objects (hereafter FUors), which are characterized by a significant increases in luminosity by orders of magnitude over decades or longer, represent such accretion outbursts. During these events, mass accretion rates can increase by three orders of magnitude \citep{Fischer:2023aa}, enabling significant mass growth within a relatively short time compared to duration of the main accretion phase \citep[$\rtsim0.5~\Myr$;][]{Evans:2009aa}. Hence, the FU Orionis-type accretion outburst may play a crucial role in the mass accretion process of low-mass stars. Several mechanisms have been proposed as an origin of the FU Ori outburts: gravitational/thermal/magnetorotational instabilities in circumstellar disks \citep{Armitage2001, Zhu2010, Bae2014}; infall of gaseous clumps formed via gravitational instability (GI) and fragmentation in disks \citep{VorobyovBasu2005, VorobyovBasu2015, 2017MeyerVorobyov}; a close encounter between a disk and an intruder star \citep{Forgan2010,Dong:2022aa,Borchert2022} or a compact cloudlet \citep{Dullemond2019,Demidova2023}; and planet-disk interaction and mass exchange \citep{Nayakshin:2012aa,Nayakshin2023}. A cascade effect involving a sequence of multiple mechanisms that eventually leads to a burst has also been suggested \citep{Skliarevskii2023}.  

Some of these mechanisms, such as GI and disk-companion interactions, predominantly operate in the outer regions of disks ($r\gtrsim10~\au$). These processes have been investigated through observations of protoplanetary disks in millimeter-wave dust thermal emission and infrared (IR) scattered light. In particular, recent observations of FUors at high angular resolutions revealed large-scale structures such as spirals and arcs extending hundreds of au in dust continuum and scattered light \citep[e.g.,][]{Liu:2016aa, Takami:2018aa, Dong:2022aa, Weber:2023ab,Weber2025}. These features suggest that GI or disk-companion interactions may trigger the accretion outburst. However, despite the success of these dust observations in several FUors, the interpretation of the dust thermal and scattered emission can still be subject to uncertainty. For example, spirals or arc-like structures seen in the IR scattered light may also originate from infalling envelopes or the trajectory of a past flyby event \citep{Weber:2023ab}. Moreover, GI-driven spirals are not always prominent in millimeter dust continuum emission \citep{Hall:2019aa}. These limitations highlight the need for alternative observational approaches to reliably identify the physical mechanisms driving FU Orionis-type outbursts.

An alternative and complementary approach is to investigate the gas kinematics of protoplanetary disks. 
\cite{Vorobyov2021} explored this possibility by analyzing the kinematic signatures of model disks during outburst phases. 
Specifically, they examined three models of the accretion and luminosity bursts: (1) activation of magnetorotational instability (MRI) in the inner disk, (2) migration and accretion of a massive gas clump (formed as a result of gravitational fragmentation of the disk) by the central star, and (3) a close encounter of two stellar objects, one of which is surrounded by a protoplanetary disk. 
This previous work demonstrated that all three models exhibit distinct kinematic features of the burst-triggering disks.

This study is a logical continuation of the work by \cite{Vorobyov2021}. In the current work, we investigate the same models to further determine their potential observational signatures, but this time with a more sophisticated analysis of the model velocity fields based on radiative transfer simulations, anticipating observations with the Atacama Large Millimeter/submillimeter Array (ALMA). The structure of this paper is as follows. Section \ref{sec:sim_hydro} provides a brief overview of the the models considered and the hydrodynamical simulation setup. Section~\ref{sec:sim_hydro-results} presents the results of the hydrodynamical simulations. Section \ref{sec:sim_obs} outlines the post-processing steps to generate synthetic molecular line data. Section \ref{sec:res_obs} discusses the results of the synthetic line observations. Section~\ref{sec:conclusion} summarizes our conclusions.


\section{Model description and simulation setup}
\label{sec:sim_hydro}
We provide a brief description of the hydrodynamical models to simulate the bursts, which are further used to construct synthetic molecular line data in later sections. The reader can find a detailed description of the numerical models of the bursts in \citet{Vorobyov2021}. The current study considers protoplanetary disks directly during the onset of the luminosity burst. These disks were obtained as a result of self-consistent modeling of the evolution of a young stellar object, starting from the collapse of the molecular cloud. The calculations are based on the Formation and Evolution Of Stars And Disks (FEOSAD) hydrodynamic model \citep{2018VorobyovAkimkin, Vorobyov2021}. During the numerical simulations, a system of hydrodynamic equations describing the dynamic evolution of the protoplanetary disk in the thin-disk approximation was numerically solved:
\begin{equation}
\label{eq:cont}
\frac{{\partial \Sigma_{\rm g} }}{{\partial t}} + \nabla \cdot
\left( \Sigma_{\rm g} {\bl v} \right) = 0,
\end{equation}

\begin{equation}
\label{eq:mom}
\frac{\partial}{\partial t} \left( \Sigma_{\rm g} {\bl v} \right) + \nabla \cdot \left( \Sigma_{\rm
g} {\bl v} \otimes {\bl v} \right) = - \nabla {\cal P} + \nabla \cdot \mathbf{\Pi} + \Sigma_{\rm g} \, {\bl g},
\end{equation}

\begin{equation}
\frac{\partial e}{\partial t} +\nabla \cdot \left( e {\bl v} \right) = -{\cal P}
(\nabla \cdot {\bl v}) -\Lambda +\Gamma +
\left(\nabla {\bl v}\right):{\bl \Pi},
\label{eq:energ}
\end{equation}
where $\Sigma_{\rm g}$ is the gas surface density, ${\bl v}$ is the gas velocity in the disk midplane, $e$ is the internal energy per unit area, ${\cal P}$ is the gas-kinetic pressure integrated over the disk thickness, $\Pi$ is the viscous stress tensor, ${\bl g}$ is the gravitational acceleration in the disk midplane, including the contribution of the central star and the disk self-gravity, $\Lambda$ is the rates of dust cooling, and $\Gamma$ is the heating of the disk by radiation from the star and external medium. The hydrodynamic equations~(\ref{eq:cont})--(\ref{eq:energ}) were solved in the polar coordinate system $\left( r, \phi \right)$ using the finite volume method, similar in methodology to the ZEUS code \citep{SN1992}. We utilize a third-order accurate, piecewise-parabolic reconstruction of fluxes \citep{Colella1984}, which guaranties low numerical diffusion. We note that FEOSAD also includes the dynamics of dust but this feature is not utilized in the current work and dust was assumed to be dynamically coupled with gas with a constant dust-to-gas mass ration of 0.01. Decoupled dust dynamics will be considered in a separate study, which focuses on distinguishing the burst mechanisms based on the dust thermal emission in mm-wavelengths. Using hydrodynamic modeling, we obtained information about the spatial distributions of gas surface density, gas velocity, gas temperature ($T_{\rm g}$), and gas disk scale height ($H_{\rm g}$)  during the luminosity outburst to be further used in the radiative transfer simulations. We note that $H_{\rm g}$ is obtained assuming the local hydrostatic equilibrium in the gravitational field of both the star and the disk \citep{VorobyovBasu2009}.

\begin{figure*}
\includegraphics[width=\linewidth]{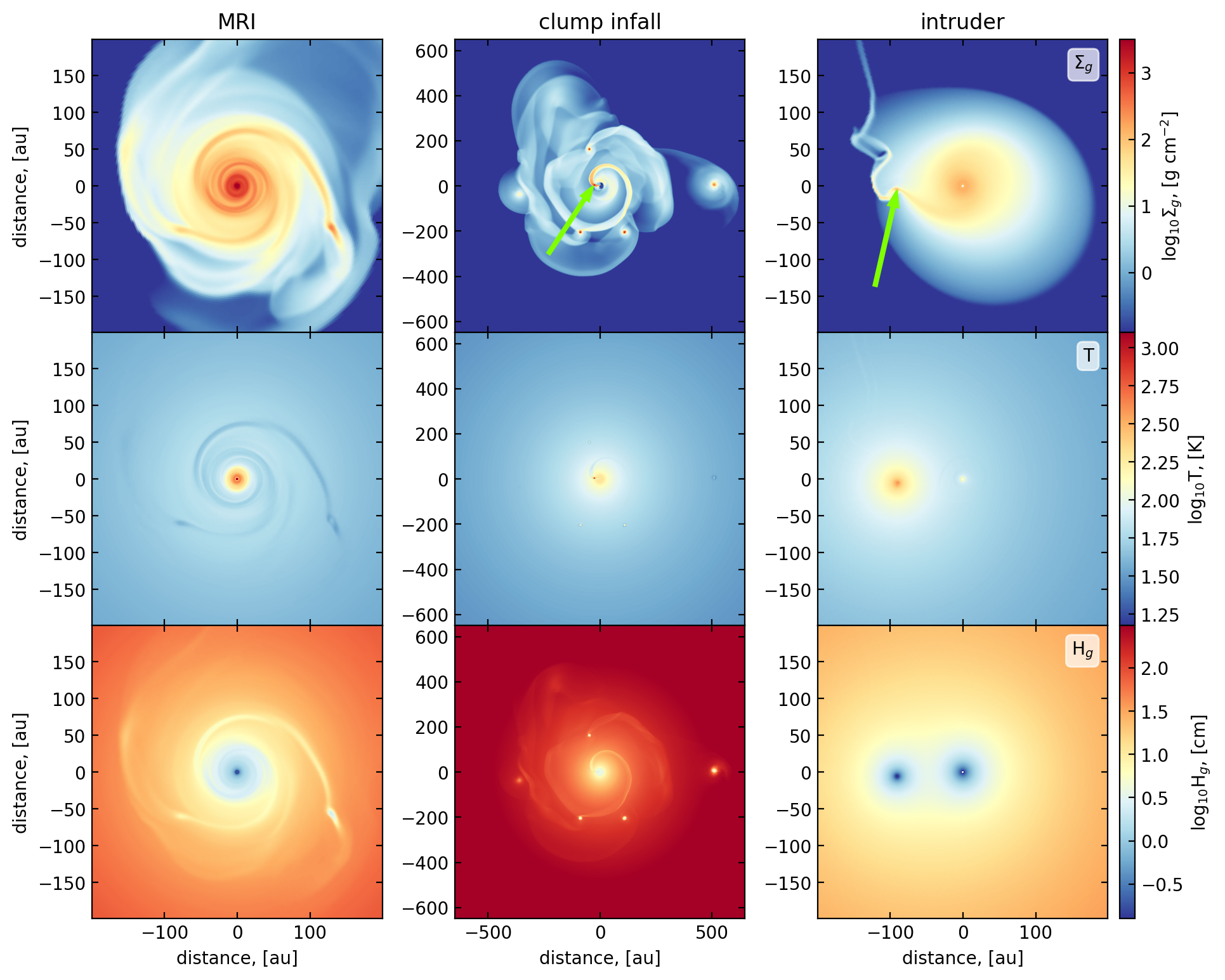}
\caption{Simulated spatial distributions of gas surface density (top row), gas temperature (middle row), and gas scale heights (bottom row) at the time of burst onset for the MRI-triggered burst (left), clump infall (center), and close stellar flyby (right).  The arrows point to the infalling clump causing the burst and the intruder. }
\label{fig:init-gas}
\end{figure*}

\section{Results of hydrodynamical simulations}
\label{sec:sim_hydro-results}

In this section, we describe the results of numerical hydrodynamic simulations of three accretion and luminosity outbursts, each having a particular burst-triggering mechanism. 

\subsection{Three outburst models}
The spatial distributions of $\Sigma_{\rm g}$, $T_{\rm g}$ (in the disk midplane), and $H_{\rm g}$  for three burst models considered are presented in Figure~\ref{fig:init-gas}. The left column presents the model of the MRI-triggered burst. In this model, the accretion and luminosity burst is caused by the MRI activation in the innermost, usually MRI-inactive disk regions, followed by a significant increase in turbulent viscosity and in the rate of mass transport onto the star across the inner regions of the disk. The MRI activation is due to exceeding the threshold temperature, at which the disk becomes sufficiently ionized (via thermal ionization of alkaline metals) to activate MRI throughout its vertical thickness, in accordance with the layered disk model of \citet{Gammie1996} and \citet{Armitage2001}. The protoplanetary disk in this model features spiral arms, which are a consequence of the development of gravitational instability. We note that gravitational instability also plays a role in the mechanism of mass pumping and subsequent temperature increase in the inner disk regions for the MRI-activated bursts \citep{Bae2014}. The midplane temperature at the positions of the spirals is lower than the environment around them. This is a direct consequence of stellar irradiation of the disk during the burst, as well as a local increase in the density and optical depth in the spirals (recall that these are density waves). This eventually leads to a time delay in heating the midplane interiors of the spiral compared to the rarefied interspiral regions at the initial stages of the burst development, an effect also reported by \citet{Laznevoi2025} in the context of axisymmetric disks. Indeed, the spatial distributions of temperature and density  shown in Figure~\ref{fig:init-gas} correspond to a time  $\approx 30$~yr after the burst onset (see Figure~\ref{fig:arates}), which may be too short to warm up the denser spirals (see Appendix~\ref{App:temp-inv} for comparison with the preburst phase). The gas scale height generally increases outward, but the spirals are of smaller scale heights than their surroundings because of the effects of lower temperature and higher disk self-gravity, which are accounted for when calculating the vertical scale height. 

We note that at the distance where spiral arms are strongest (several tens of au) the dominant source of heating is stellar irradiation, while viscous heating dominates only in the inner disk because the viscous heating rate declines with distance faster than that of radiative heating (see Fig.~9 in \citealt{2018VorobyovAkimkin}). The PdV work (adiabatic heating/cooling; first right-hand term in Eq.~\ref{eq:energ}) can also be an important source of heating within spiral density waves and self-gravitating clumps. All three heating mechanisms are self-consistently considered in our numerical model (see Equation~\ref{eq:energ}).

The next considered mechanism of the luminosity outburst, described in~\cite{VorobyovBasu2005, Machida2011,2017MeyerVorobyov}, is presented in the middle column of Figure~\ref{fig:init-gas}. In this case, the burst is triggered due to the tidal destruction and accretion of a massive gaseous clump migrating from the outer regions of the disk towards the star. The gravitationally unstable disk in this model has a significantly larger radial extent, which makes disk gravitational fragmentation easier \citep{VorobyovBasu2015}.  
Five clumps are localized at distances $\ge 200$~au from the star and one more is at the head of a strong one-armed spiral in close proximity to the central star. The gravitational energy of this closest clump is converted into thermal energy and released in the form of radiation from the star during its tidal destruction and accretion onto the star \citep{2018VorobyovElbakyan}. We note that the clumps are warmer than the surrounding disk due to high optical depths and adiabatic heating by clump's self-gravity. The temperature inversion in the midplane of the one-armed spiral is much less pronounced than in the case of MRI-triggered burst, because this spiral consists of already warm material that is tidally stripped off the inward migrating clump. The clumps are of small gas scale heights because of strong self-gravity. ALMA observations of V960 Mon revealed multiple fragments of 1.3 mm dust continuum emission aligned along a spiral shape, which may suggest such gravitationally unstable spirals that fragment into massive clumps.

Finally, the third model, shown in the right column of Figure~\ref{fig:init-gas}, is a system of two interacting stars, one of which is surrounded by a protoplanetary disk, and the other diskless star (hereafter, the intruder) makes a prograde approach relative to the central star. This is the so-called stellar ``collision'' or ``close flyby'' model. This model is distinguished by the fact that the luminosity outburst there occurs not on the central star, but on the intruder. In this case, the increased luminosity is a consequence of the natural increase in the amount of accreted matter due to the motion of the intruder relative to the disk of the central star. Morphologically, the system under consideration is distinguished by the presence of a characteristic comet-like structure along the trajectory of the intruder. Moreover, we can clearly notice the characteristic dipole distribution in the disk scale height. 

The velocity of the intruder at periastron is $\approx 5$~km~s$^{-1}$ and declines on both sides of the intruder's trajectory. This may not be sufficient to explain the timing of the outburst in FU~Ori, given the current separation between the companions of $\approx$ 250~au \citep{Liu2017} and the fast-rising light curve (which requires a periastron of $\le 10$~au; \citealt{Borchert2021a}). A possible solution was proposed by \citet{Skliarevskii2023} who introduced a cascade outburst scenario. A distant and slow stellar flyby with a periastron of a few hundred astronomical units exerts a gravitational perturbation to the disk of the target, which is sufficient to drive the inner disk out of marginal equilibrium and trigger thermal and magnetorotational instabilities one after another.
In this scenario, the target star is bursting, rather than the intruder, and its kinematic signatures deserve a separate study. Our case is relevant to  outbursts caused by  a penetrating collision, having the kinematic characteristics expected for a highly perturbed disk. It should also be noted that, as a result of the restrictions imposed by the thin-disk model, the approach of the two objects occurs in the plane of the disk of the central star. Close approaches with an arbitrary trajectory and the bursts caused by them are considered, for example, in \cite{Cuello2023} and \cite{Demidova2023}.

\begin{figure}
\centering
\includegraphics[width=\linewidth]{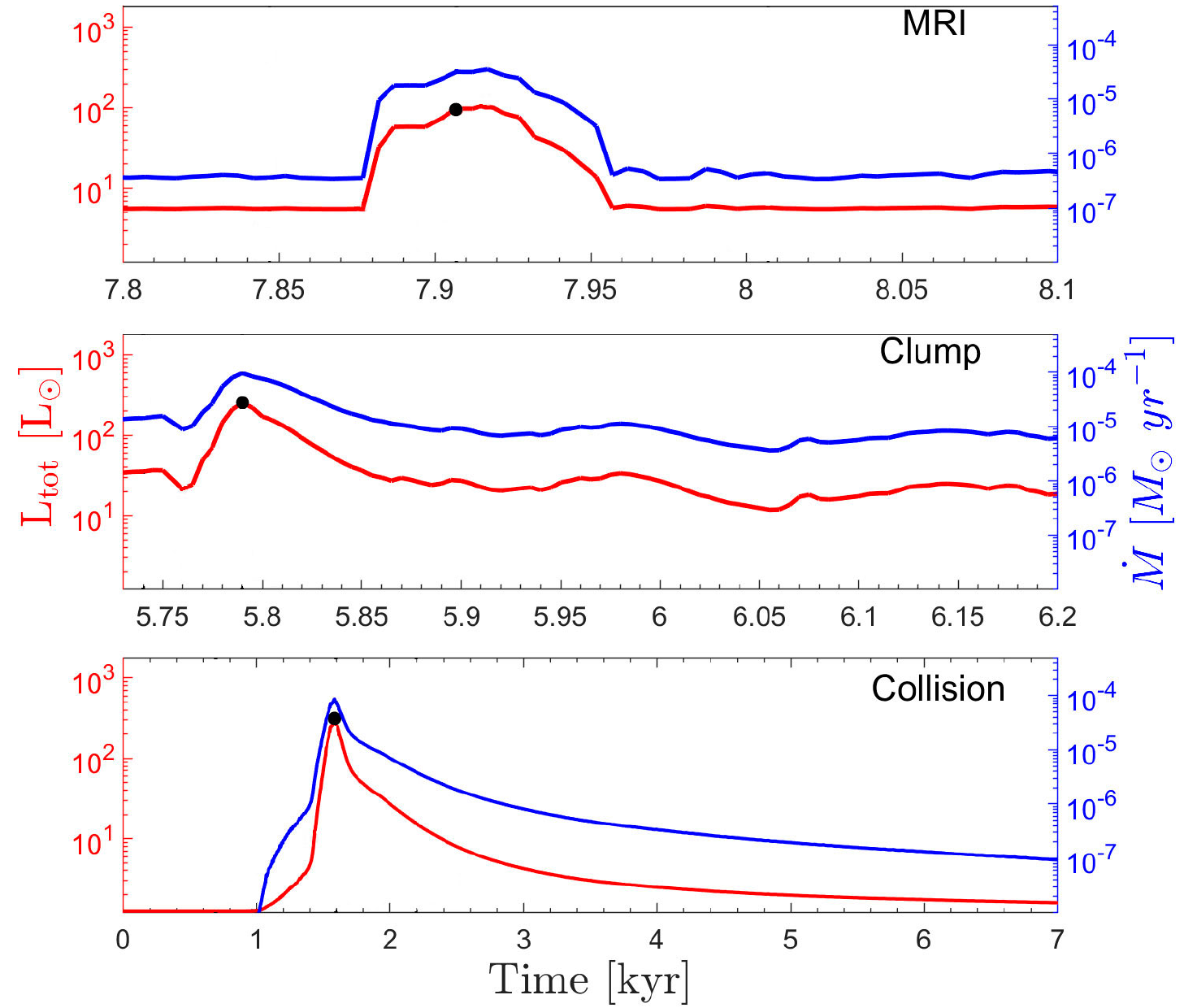}
\caption{Mass accretion rates and total luminosities in the three burst models considered. The panels from top to bottom correspond to the MRI-triggered burst, clump accretion, and close flyby, respectively. The black circles indicate the exact times during the bursts considered in this work.   }
\label{fig:arates}
\end{figure} 

Figure~\ref{fig:arates} depicts the evolution of the mass accretion rates and total luminosities in a time period encompassing the burst in the three models considered. The mass accretion rate $\dot{M}$ is calculated as the mass passing through the inner computational boundary per unit time, while the total luminosity is the sum of accretion luminosity ($0.5 G M_\ast \dot{M} / R_\ast$) and photospheric luminosity, where $G$ is the gravitational constant, $M_\ast$ the stellar mass, and $R_\ast$ the stellar radius. The stellar photospheric luminosity and stellar radius are taken from the precomputed stellar evolution tracks of \citet{2008YorkeBodenheimer}. The peak luminosities of model bursts are in the range of the known FU Orionis-type bursts \citep[see Table A.1 in][]{Vorobyov2021}. The burst duration in the MRI-triggered and clump-infall models are within a hundred years, also consistent with FU Orionis objects. The burst duration in the close encounter model is longer, which is explained by the in-plane character of simulated flyby. 
The main parameters of the models and the bursts are listed in Table~\ref{tbl:initial}.

\begin{deluxetable}{lcccc}
\tablewidth{\textwidth}
\tablecaption{Main parameters of the models under consideration.}
\tablehead{\colhead{Model} & \colhead{$M_{\rm disk}$} & \colhead{$M_{\ast}$} & \colhead{$L_{\ast}$}  & \colhead{$r_{\rm per}$} \\
\colhead{} & \colhead{$(M_{\odot})$} & \colhead{$(M_{\odot})$} & \colhead{$(L_{\odot})$}  & \colhead{$(\au)$}
}
\colnumbers
\startdata
MRI-triggered  & 0.36 & 0.63 & 104 & -- \\ 
Clump infall & 0.19 & 0.55 & 254 & --\\ 
Flyby (target) & 0.06 & 0.47 & 1.6 & 82.5 \\ 
Flyby (intruder) & - & 0.5 & 273 & 82.5 \\ 
\hline
\enddata
\tablecomments{(1) Model name. (2) Mass of the disk surrounding the star. (3) Mass of the star. (4) Luminosity of the star during the peak of the outburst. (5) Periastron distance for the flyby model.}
\label{tbl:initial}
\end{deluxetable}

\begin{figure*}
\includegraphics[width=\textwidth]{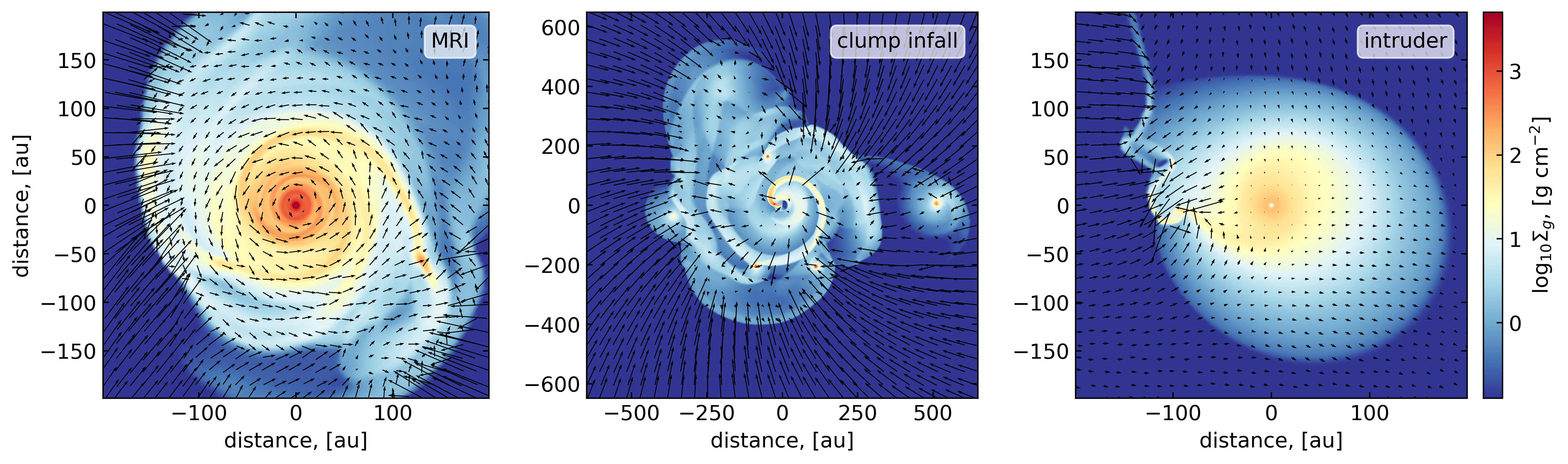}
\caption{Surface density (color) and velocity field (vectors) in the MRI model (left), clump-infall model (middle), and close stellar flyby model (right). The Keplerian rotation is subtracted from the model velocity. Note the different spatial scales to better resolve the disk spatial morphology and velocity fields. The disk rotates counterclockwise. The scale bar is in g~cm$^{-2}$.}
\label{fig:quiver}
\end{figure*}

\subsection{Model velocity fields}
Figure~\ref{fig:quiver} presents the gas surface densities and residual velocities (after subtracting the Keplerian velocity) in all three models. 
In the MRI model, the rotation is nearly Keplerian in the inner 10~au and is super-Keplerian in the intermediate and outer disk regions because of the contribution from the disk self-gravity to the gravity of the star. Since the disk is still accreting from the remnant of the parental cloud core, the velocity pattern outside the disk is characterized by strong infall motion. The gravity force from the non-axisymmetric spiral arms perturbs the gas as it rotates around the star, deflecting the gas flow from a purely circular motion. The gas is accelerated as it approaches from behind the spiral arms. The gas is then caught by the gravitational field of the spiral arm and moves along the spiral arm for some time. This later effect is clearly seen from the velocity vectors being aligned with the spiral density enhancements. Finally, the gas leaves the spiral arm and resumes its near-circular motion towards the next spiral arm, thus repeating the velocity pattern described above.


In the clump-infall model of the burst, the velocity pattern outside the disk also shows a notable inflow motion attributed to the infalling envelope, the mass loading from which promotes disk gravitational fragmentation. However, the velocity pattern inside the disk is conspicuously distinct from the MRI-triggered case. Strong outflow motions are caused by angular momentum redistribution due to the gravitational interaction between the infalling clump and the spiral arm. The former loses angular momentum and migrates towards the star, causing a burst when tidally destroyed, while the latter gains angular momentum and expands outward.\footnote{An animation of this process can be found at \url{https://jinshisai.github.io/animations/accretion_clip.mp4}.} Strong deviations from the Keplerian motion in the vicinity of the spiral arm are expected in this model.  

In the flyby model, the penetrating collision with an intruder star produces a strong and rather complex perturbation to the velocity field of the target disk in the vicinity of the intruder. The disk material is drawn towards the intruder in the nearest side of the disk, 
while the opposite side of the disk expands due to tidal forces exerted on the star+disk system by the intruder. Strong non-Keplerian motion is expected in this case throughout the disk.

\section{Synthetic Observations} \label{sec:sim_obs}

To simulate molecular line observations of the above three numerical models with ALMA, we performed radiative transfer calculations using \texttt{RADMC-3D}\footnote{\url{https://www.ita.uni-heidelberg.de/~dullemond/software/radmc-3d/}}. The volume distributions of the gas density, temperature, and velocity in the considered disk models were reconstructed using an assumption of the vertical hydrostatic equilibrium and the information on the gas scale height (see Figure \ref{fig:init-gas}). Dust grains were assumed to be fully coupled with gas; therefore the gas scale height was also adopted for dust. We neglected possible variations of the gas velocity and temperature with disk height and set these quantities equal to those in the disk midplane.

We simulated observations in for the \ce{C^18O} $J=3$--2 (329.33055250 GHz) line. \ce{C^18O} is an optically thin disk tracer commonly used to study young protostellar disks embedded within infalling envelopes \citep[e.g.,][]{Ohashi:2023aa}. 
Due to its relatively high excitation energy, the \ce{C^18O} $J=3\mbox{--}2$ transition is expected to offer good transparency through envelopes and parental clouds, making it well-suited for probing gas kinematics of embedded disks. We note that our hydrodynamical simulations are in two dimensions ($r,\phi$) and do not include vertical envelope infall. Thus, there is no envelope contamination along the line of sight when a moderate inclination angle is assumed. The possibility of envelope contamination along the line of sight is not examined in this work since it requires detailed modeling of envelope structures and chemistry, which is beyond the scope of this paper. 

We adopted a \ce{C^18O} abundance with respect to \ce{H_2} gas of $1.7 \times 10^{-7}$ \citep{Frerking:1982aa}. The background dust continuum emission was calculated with a dust opacity table from DSHAPR \citep[][]{Birnstiel:2018aa} and subtracted from the \ce{C^18O} emission. Only dust absorption opacity was considered, and dust scattering was not included. For the synthetic observations, we assumed an inclination angle of $i=30^\circ$, a position angle of $90^\circ$ and a velocity channel width of $0.16~\kmps$. The model protostellar systems are configured such that the far and near sides of the disk correspond to the northern and southern sides of the protostar on the plane of the sky, respectively. The source distance was set to $300~\pc$, which can be considered as an average distance to a group of closest FU~Orionis and FU~Orionis-like objects \citep{Audard:2014aa,Connelley:2018aa,Vorobyov2021}.

We further post-processed the synthetic \ce{C^18O} images by adding random noise and convolving a circular Gaussian beam, assuming a rms noise of $2.5~\mjypbm$ (corresponding to 2.8 K) and a beam full-width-half-maximum (FWHM) of $0\farcs1$, which are close to ALMA's sensitivity limits but still achievable. We did not simulate full ALMA observation simulations, such as $uv$-sampling and image reconstruction via the CLEAN algorithm, as our goal was not to reproduce a specific ALMA data set. Previous observational studies have shown that ALMA imaging fidelity is sufficient to probe detailed gas kinematics of protoplanetary disks and detect small velocity deviations of a few percent of Keplerian velocity, especially for nearby sources \citep[e.g.,][]{Teague:2018aa, Pinte:2018aa, Teague:2025aa, Izquierdo:2025aa}. Hereafter, we refer to the synthetic data without and with beam convolution as the fiducial and beam-convolved data, respectively.

\section{Results of Synthetic Observations} \label{sec:res_obs}

In this section, we present the results of synthetic observations of the three models in the \ce{C^18O} $J=3\mbox{--}2$ line. Section~\ref{subsec:morphology} discusses the morphology of the line emission, followed by an analysis of gas kinematics in Section~\ref{subsec:kinematics}.

\subsection{Morphology of line emission}\label{subsec:morphology}

\begin{figure*}[htbp]
\centering
\includegraphics[width=\textwidth]{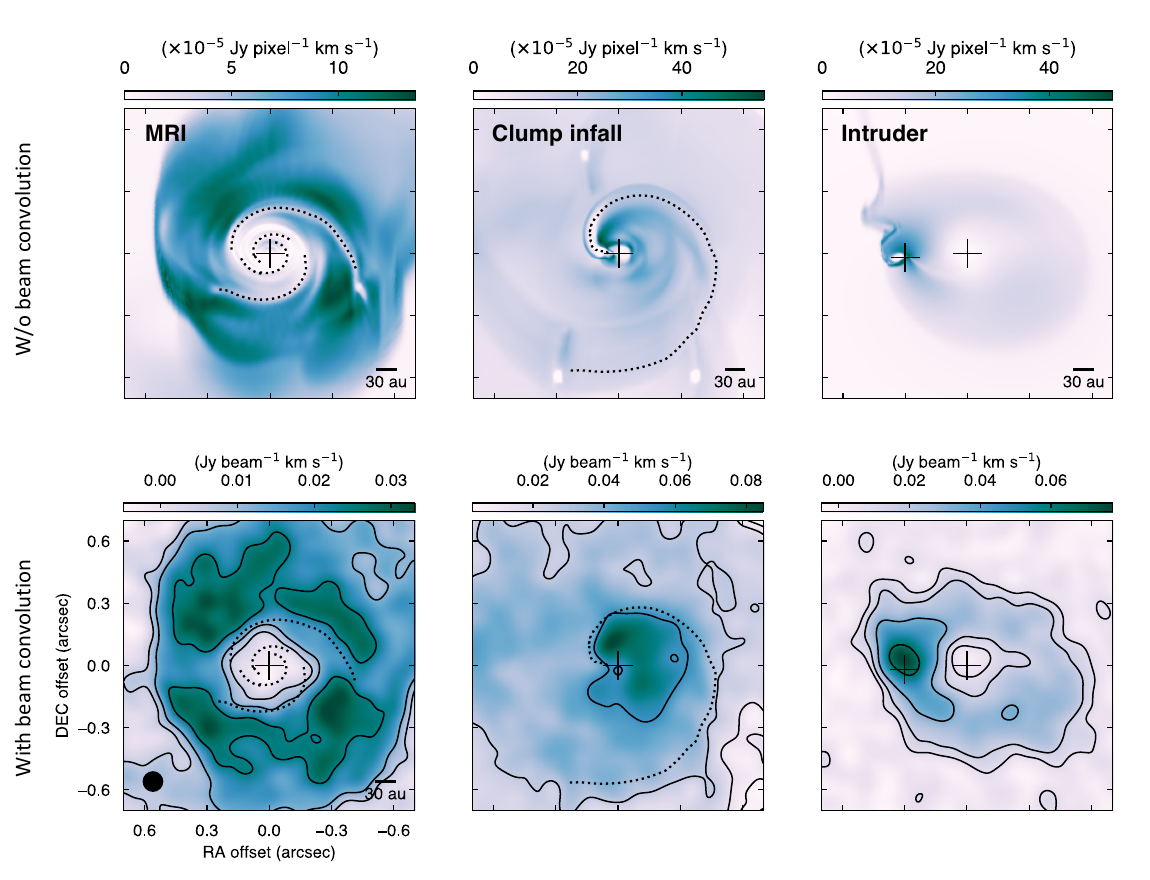}
\caption{Integrated intensity maps of the synthetic \ce{C^18O} $J=3\mbox{--}2$ emission for the three models without (top row) and with (bottom row) beam convolution. From left to right, maps for the MRI, clump-infall, and intruder models are presented. The dotted curves indicate the locations of the high-density spiral arms in the first two models. The crosses mark positions of the central protostars and the intruder. In the bottom row, contour levels are $3$, 6, 12, and $24\times\sigma$, where $\sigma = 2.0$, 2.2, and $2.6~\mjypbm$ for the MRI, clump-infall, and the intruder models, respectively. The filled circle in the bottom left panel denotes the beam size of $0\farcs1$.}
\label{fig:mom0s}
\end{figure*}

The top row of Figure \ref{fig:mom0s} shows the integrated intensity maps of the fiducial synthetic \ce{C^18O} $J=3$--2 data of all models without beam convolution and noise. In the MRI model, the integrated \ce{C^18O} emission exhibits spiral-like morphologies. However, they do not coincide with the actual locations of the high-density spiral arms, which are denoted by dotted curves. Furthermore, the spiral arms appear weaker in intensity than their surroundings. This discrepancy arises because in the disk midplane the spiral arms are colder than the surrounding gas during the outburst due to their higher densities and optical depths (see Appendix \ref{App:temp-inv}). The \ce{C^18O} emission is absent around the center in the integrated intensity map. This absence of the \ce{C^18O} emission is likely due to the dust continuum becoming optically thick at small radii and reaching a brightness comparable to the line emission, resulting in the \ce{C^18O} emission being removed by continuum subtraction.

The clump-infall model shows similar characteristics. The integrated emission is absent at the inner regions of the spiral arm, where the clump is being tidally dispersed, and around the protostar due to the high optical depth of the dust continuum emission. Whereas the intensity peaks appear to be in spiral-like morphologies just ahead of and behind the actual location of the high-density outer spiral arm, the emission along the spiral itself is weaker than or comparable to that of the surrounding gas. This is also attributed to the lower temperature of the spiral arm. These findings indicate that line intensity is not a reliable tracer of the total gas density alone, since it is also sensitive to gas temperature. The time delay in warming the denser spirals compared to rarefied inter-arm regions leads to a peculiar temperature distribution during the outburst, which complicates the interpretation of line intensity maps.

In the intruder model, the integrated intensity of the \ce{C^18O} emission is strongest around the intruder primarily due to elevated gas temperature resulting from the accretion and luminosity outbursts. This produces a pronounced asymmetry in the integrated intensity map. 
Additionally, the \ce{C^18O} emission traces a comet-like structure along the intruder's trajectory.

The bottom row of Figure \ref{fig:mom0s} presents integrated-intensity maps for all models after convolution with a Gaussian beam with FWHM of $0\farcs1$ (corresponding to $\rtsim30~\au$) and the addition of random noise of $2.5~\mjypbm$. After beam convolution, the morphology of the line emission diverges even further from the original gas density distribution. In both the MRI and clump-infall models, the intensity peaks do not coincide with the spiral structures, as was the case prior to beam convolution. Moreover, in the MRI model, the emission appears more ring-like than spiral-shaped. As the central hole making the emission ring-like shape is due to the optically thick dust continuum emission, it may be difficult to differentiate the MRI model from symmetric disks that are not under GI but have optically thick dust continuum emission near the center based solely on the line morphology. In the clump-infall model, the emission is asymmetric and does not show spiral-like shapes. The intensity peak does not correspond to any of the remaining clumps in the disk but instead traces warm, optically thin material located behind the spiral arm. In the intruder model, the integrated emission shows a peak near the intruder's position and a central hole caused by the optically thick dust continuum emission, resulting in a ring-like structure with a crescent-shaped asymmetry.

These results highlight that the morphology of line emission can differ significantly from the underlying gas density structures in outbursting disks and must be interpreted with caution. On the other hand, the asymmetric intensity distributions in the clump-infall and intruder models, caused by asymmetric temperature structures, are distinct from the MRI model. Therefore, these features may help differentiate the models when combined with the kinematic diagnostics discussed in the following section. We note that the observed intensity distribution depends on the choice of molecular tracer and transition, and may vary for other lines. 

\subsection{Kinematics of line emission}\label{subsec:kinematics}


\begin{figure*}[tbhp]
\centering
\includegraphics[width=0.95\textwidth]{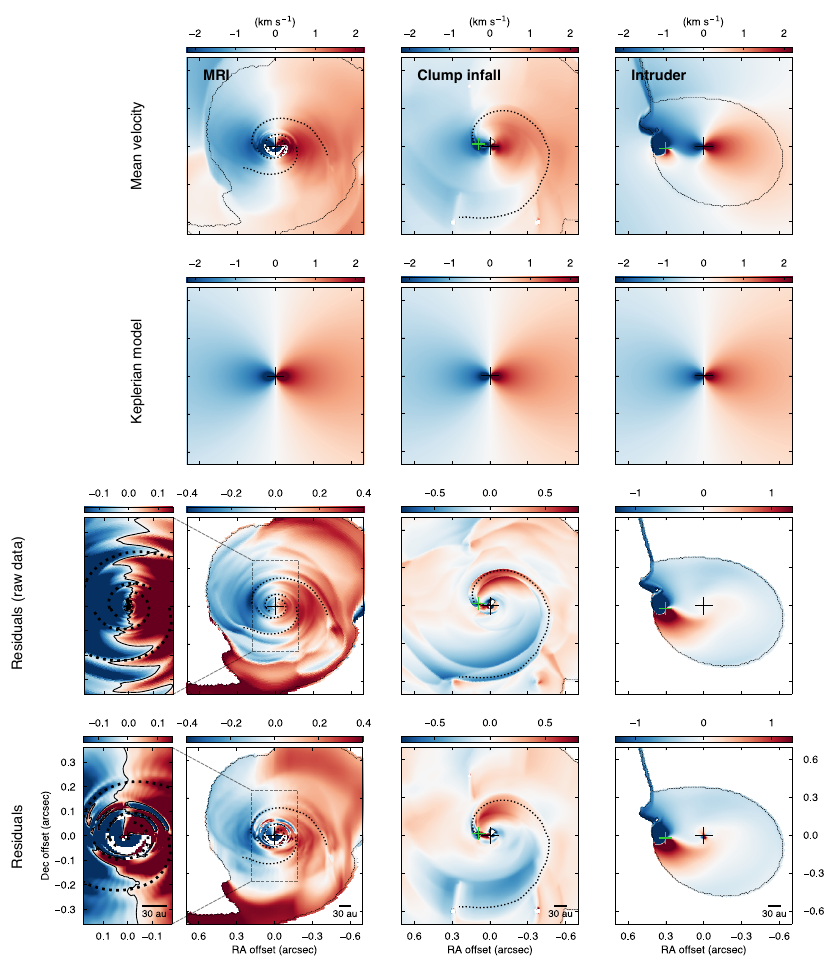}
\caption{Mean-velocity maps of the synthetic \ce{C^18O} emission (first row); expected Keplerian velocity fields after projection onto the plane of the sky (second row); residual-velocity maps of the raw simulation data, which are computed using the true gas velocity from the simulations (third row); and residual-velocity maps of the synthetic \ce{C^18O} emission (forth row). From left to right, maps for the MRI, clump-infall and intruder models are presented. We note that color scales of the residual maps are different for all models. The boundaries between the disks and envelopes, which are defined by a density threshold of $0.1~\mathrm{g~cm^{-2}}$, are denoted by contours. The envelope regions are masked in the residual maps for better visualizations. The dotted curves indicate the locations of the high-density spiral arms. Black crosses indicate the protostellar positions, and green crosses denote positions of an inner clump or the intruder. Contours in the zoom-in view of the residual maps for the MRI model denote zero residual velocities.}
\label{fig:mom1_all}
\end{figure*}

\begin{figure}[tbhp]
\centering
\includegraphics[width=0.5\textwidth]{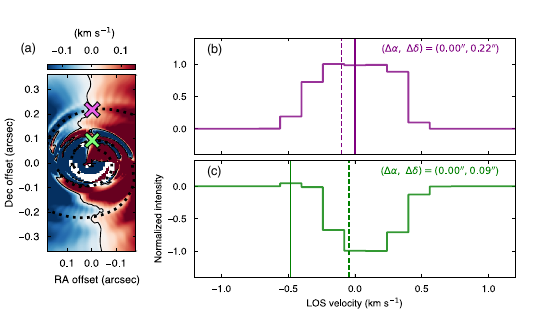}
\caption{(a) Same as the zoom-in view of the residual map of the MRI model in Figure \ref{fig:mom1_all} but with markers that denote the positions where the spectra shown in the right panels were measured. (b, c) Spectra of the synthetic \ce{C^18O} emission of the MRI model, measured at the positions of the northern, inner and outer spirals around the disk minor axis. Vertical solid and dashed lines indicate intensity-weighted mean velocity calculated with the spectra and the true gas velocity at the positions in the numerical simulation, respectively.}
\label{fig:spec_mri_nobeam}
\end{figure}

\begin{figure*}
\centering
\includegraphics[width=0.7\textwidth]{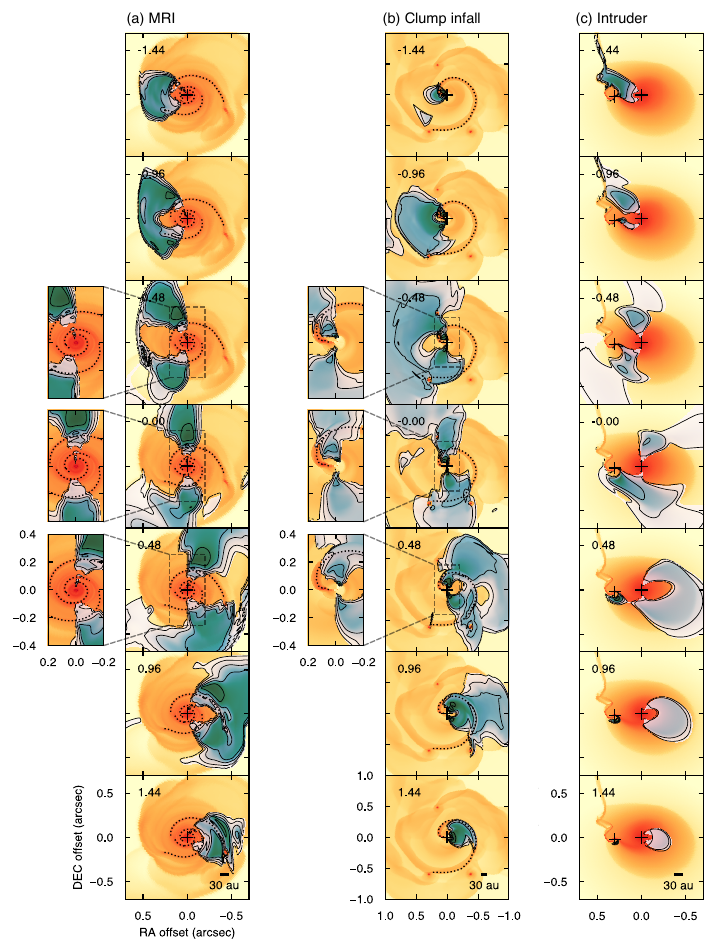}
\caption{Velocity channel maps of the synthetic $\CeighteenO$ $J=3$--2 emission of the three models without beam convolution. The channel maps are presented in steps of $0.48~\kmps$, while the original velocity resolution of the synthetic data is $0.16~\kmps$. Green color and black contours show the synthetic $\CeighteenO$ emission. Contour levels are 3, 10, 30, 60, and 90\% of the maximum intensity. The background images in orange show the surface density distributions of the input models. We note that the map extent of the channel maps for the clump-infall model is larger than those for the other two models. The dotted curves highlight the locations of the high-density spiral arms.}
\label{fig:channel_all}
\end{figure*}

In this section, we examine the kinematic features revealed in the \ce{C^18O} emission. 
Given the fixed geometry of the disk, the observed line-of-sight (LOS) velocity can be decomposed into radial, azimuthal, and vertical velocity components of the disk. 
In the current work, the projection of cylindrical coordinates $(r, \phi, z)$ on the disk frame onto the plane of the sky is defined as:
\begin{align}
    &\Delta \alpha = - x, \\
    &\Delta \delta = y \cos i + z \sin i,
\end{align}
where $i$ is the inclination angle of the disk, $\Delta \alpha$ and $\Delta \delta$ are the right ascension and declination offsets, respectively. Cartesian coordinates $x$ and $y$ are given by:
\begin{align}
    &x = r \cos\phi, \\
    &y = r \sin\phi.
\end{align}
The observed LOS velocity is expressed as:
\begin{eqnarray}
    \vlos = v_r \sin i \sin \phi + v_\phi \sin i \cos \phi \nonumber \\
    + v_z \cos i + \vsys,
\end{eqnarray}
where $\vsys$ is the systemic velocity. In the current models, vertical motion and systemic velocity are neglected (i.e., $v_z = 0$ and $\vsys = 0$). Hence, according to the above relationships, rotational and radial velocity components are primarily observed along the disk major axis (where $\phi=0$ or $\pi$, and $\Delta \delta = 0$) and minor (where $\phi=\pi/2$ or $3\pi/2$, and $\Delta \alpha = 0$) axis, respectively.


\subsubsection{Fiducial mean-velocity maps}

The first and second rows of Figure \ref{fig:mom1_all} present the mean-velocity maps of the synthesized \ce{C^18O} emission without beam convolution and LOS velocity fields expected from pure Keplerian rotation. The projected Keplerian velocity fields are calculated with the protostellar masses listed in Table \ref{tbl:initial} and an inclination angle of $30^\circ$. The boundaries between the disks and envelopes, defined using a surface density threshold of $0.1~\mathrm{g~cm^{-2}}$, are also shown in the figures. We adopt this density-based criterion rather than a kinematical definition to define the boundaries, since it well separates the disks and envelopes in all three models without being affected by strong velocity perturbations within the disks. The mean-velocity map of the MRI model exhibits a clear feature of a rotational motion, which is very similar to the Keplerian velocity field, while radial infall motion is noticeable outside the disk. The clump-infall model also shows rotation motion of the disk, but with noticeable deviations from pure Keplerian rotation. Along the disk minor axis, the mean velocity is redshifted on the northern side and blueshifted on the southern side of the protostar, while no LOS velocity is expected from pure Keplerian rotation without radial velocity components. The intruder model exhibits strong perturbations around the intruder star, in addition to the rotation motion of the primary disk.

To better visualize deviations from Keplerian motion within the disks, we computed residual velocity maps by subtracting the projected Keplerian velocity fields from the mean-velocity maps. For comparison, we also calculated LOS velocity fields using the true gas velocities in the disk midplane from the simulations. Residual velocity maps for the raw simulation data were then derived by subtracting the projected Keplerian velocity fields from these LOS velocity fields. The residual velocity maps of the raw simulation data and the synthetic \ce{C^18O} emission are presented in the third and fourth rows of Figure \ref{fig:mom1_all}, respectively. For better visualization, envelope regions are masked in these residual maps.

In the MRI model, residuals greater than $0.1~\kmps$ are present across the disk in the residual velocity map of the raw simulation data (third row of Figure \ref{fig:mom1_all}). These residuals are blueshifted on the eastern side and redshifted on the western side of the protostar, indicating a rotational motion. This remaining rotational motion is due to the self-gravity of the disk, which causes the disk to rotate faster than it would under the star's gravitational field alone. In addition to this global trend, the residual map shows local velocity perturbations of $\rtsim0.1~\kmps$ on scales of $\rtsim10~\au$, associated with the spirals. Along the disk minor axis, residual velocities just ahead of and behind the northern outer spiral are redshifted, while the velocity at the location of the spiral is blueshifted. Similarly, blueshifted residuals are seen around the southern outer spiral with near-zero or slightly redshifted velocity at the spiral itself. Since LOS velocity along the disk minor axis reflects radial motion, these repeated redshifted and blueshifted velocities suggest gas motion moving radially inward and outward in the disk plane. In other words, the gas surrounding the spirals approaches or moves away from the spiral arms, while the gas within the spirals moves along them (see Figure \ref{fig:quiver}). The same features, albeit less pronounced, are also seen around the inner spirals. Similar velocity patterns with repeated redshifted and blueshifted residual velocities near the disk minor axis have been also reported in a previous study of a gravitationally unstable disk that is unrelated to the FU~Orionis phenomenon \citep{Hall:2020aa}.

These local velocity perturbations associated with the GI-induced spirals are less prominent in the residual maps of the \ce{C^18O} emission (fourth row of Figure \ref{fig:mom1_all}) than in the raw simulation data. This is likely due to smoothing effects caused by the disk's vertical thickness (see also Figure \ref{fig:mom1_mri_thin} in Appendix \ref{app:mri_thin}). In inclined, geometrically thick disks, gas at different disk heights and radii contributes to the same line of sight, effectively smoothing velocity structures in the direction of the disk minor axis (north-south direction in the current case). A spectrum extracted from the northern outer spiral along the disk minor axis (Figure \ref{fig:spec_mri_nobeam}b) indeed shows a skew toward redshifted velocities compared to the projected, true midplane velocity, which is consistent with contamination from nearby redshifted velocity components.

Additionally, the residual velocity map of the \ce{C^18O} emission shows a flip between blueshifted and redshifted velocities just above the northern inner spiral, which is not seen in the residual map for the raw simulation data. These complex velocity structures at small radii are likely caused by self-absorption in the \ce{C^18O} emission, as negative emission is present in the corresponding spectrum (Figure \ref{fig:spec_mri_nobeam}c).

In the clump-infall model, the residual maps of both the raw simulation data and \ce{C^18O} emission show redshifted and blueshifted residuals of $\rtsim0.5~\kmps$ on the northern and southern sides of the disk (corresponding to the far and near sides, respectively) along the spiral arm. These residuals indicate gas expansion in the vicinity of the spiral arm (see also Figure \ref{fig:quiver}). This feature of gas expansion is present across several tens au, and results from the gravitational exchange of angular momentum between the infalling clump and the spiral arm. The former loses angular momentum and migrates inward toward the star, while the latter gains angular momentum and expands outward. The counterclockwise rotation of the clump, attributed to a prior clump-clump merger event \citep{2018VorobyovElbakyan} is also noticeable. These features are more pronounced than those of the GI-induced velocity structures seen in the MRI model, both in amplitude and spatial scale. Hence, they remain clearly visible in the synthetic \ce{C^18O} emission even after the radiative transfer processing, despite the smoothing effects induced by the disk's vertical thickness.

In the intruder model, the residual velocity maps of both the raw simulation data and \ce{C^18O} emission exhibit strong perturbations around the intruder star, with weaker distortions on the opposite side of the disk due to tidal interactions with the intruder's gravitational field. The residual velocity is blueshifted on the northern side and redshifted on the southern side of the intruder, forming a velocity gradient that reflects accretion of gas onto the intruder, which are consistent with the numerical simulation. The residual velocity structures of the \ce{C^18O} emission closely match those in the raw simulation data, as their amplitude and spatial extent are large enough to be unaffected by the disk thickness.

\subsubsection{Fiducial velocity channel maps}

Further details of velocity structures are provided in velocity channel maps. Figure \ref{fig:channel_all} presents velocity channel maps of the fiducial synthetic \ce{C^18O} emission for all three models. In general, the \ce{C^18O} emission follows Keplerian rotation patterns, but each model exhibits varying degrees of perturbation.

In the MRI model, the emission shows kinks around the two outer spiral arms, particularly in the channels centered at velocities $\pm 0.48$~km~s$^{-1}$ (Figure \ref{fig:channel_all}a). These kinks are also found in velocity channel maps without radiative transfer calculations in the first paper \citep{Vorobyov2021}. Similar kinks, referred to as GI wiggles, have been reported in a previous work of an gravitationally unstable disk unrelated to FU Orionis outbursts \citep{Hall:2020aa}. These GI wiggles are responsible for the repeated blueshifted and redshifted velocities near the disk minor axis in the residual velocity map. These features arise because non-Keplerian motions around the spirals shift the line emission into adjacent velocity channels, rather than the channel corresponding to the local Keplerian velocity. Weaker wiggles are also observed around less prominent spiral structures at larger radii. Strong deviations from Keplerian motion are also evident around the disk-infalling envelope transition zone (e.g., bottom-left corner of the map). The absence of emission near the disk center is due to the optically thick dust continuum emission.

In the clump-infall model, the \ce{C^18O} emission shows similar but more pronounced kinks around the inner spiral arm in the north at velocities $\pm 0.48$~km~s$^{-1}$ (Figure \ref{fig:channel_all}b), which are consistent with velocity channel maps presented in the first paper \citep[][]{Vorobyov2021}. These kinks are caused by the same mechanism as the GI wiggles but are amplified by radial expansion due to the exchange of angular momentum between the inward-migrating clump and the spiral.

In the intruder model, the emission exhibits highly distorted velocity structures that do not follow Keplerian velocity patterns on the eastern side of the primary protostar (Figure \ref{fig:channel_all}c). At velocity of $-1.44~\kmps$, the emission is concentrated on the northern side of the intruder and appears connected to it, while the emission shifts to the southern side at velocities of $0.00\mbox{--}1.44~\kmps$. This velocity gradient is consistent with gas accreted by the intruder from north and south in the numerical simulation (see Figure \ref{fig:quiver}). On the west side of the primary protostar, the Keplerian velocity pattern is better preserved, although minor perturbations are still present.


\begin{figure*}
\includegraphics[width=\textwidth]{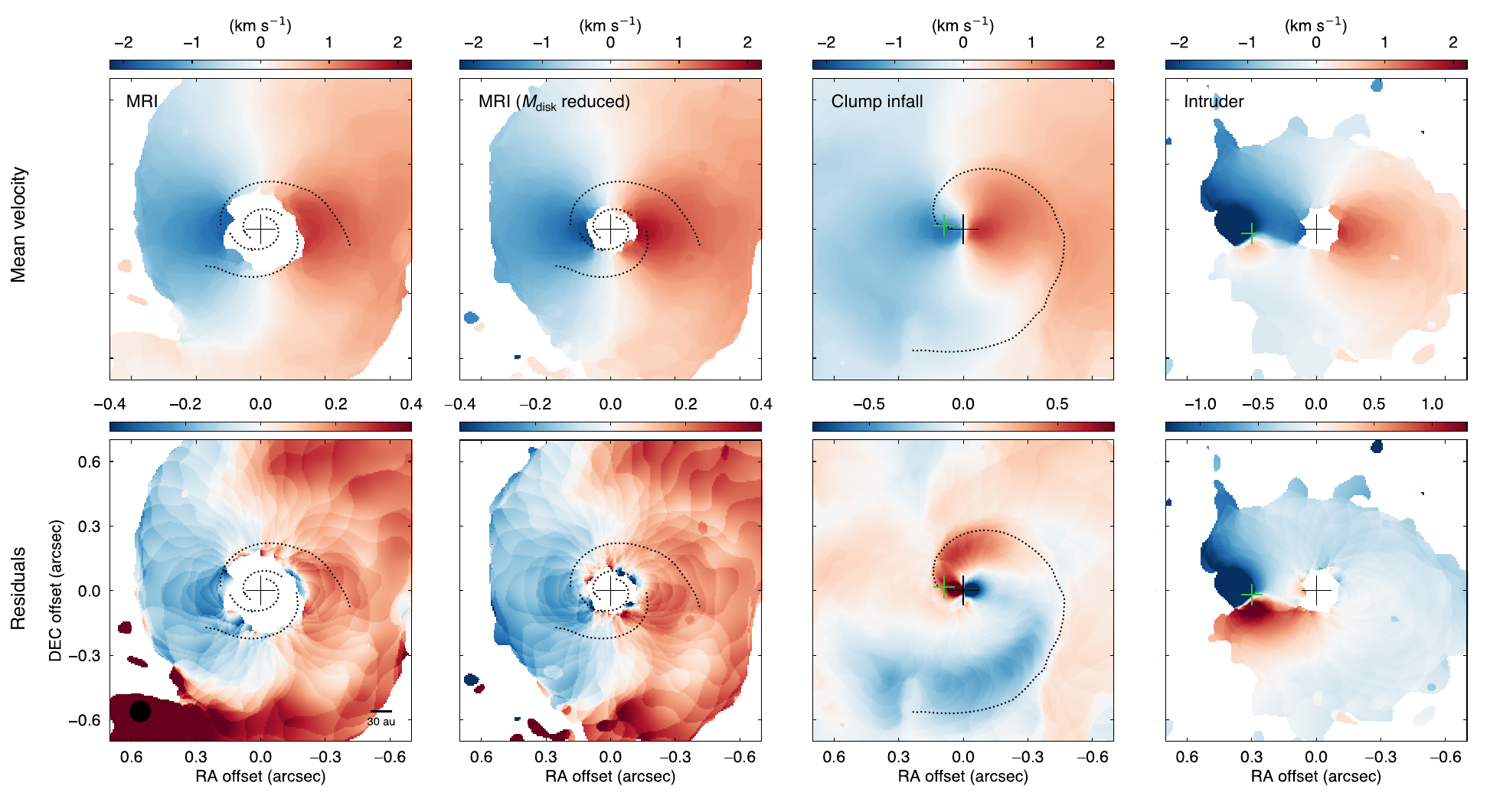}
\caption{Mean-velocity (top row) and residual (bottom row) maps of the \ce{C^18O} $J=3\mbox{--}2$ emission after beam convolution. The dotted curves indicate the locations of the high-density spiral arms. The filled circle in the bottom left panel shows the beam size of $0\farcs1$. Black crosses indicate the protostellar positions, and green crosses denote positions of an inner clump or the intruder.}
\label{fig:mom1_beam}
\end{figure*}

\begin{figure*}
\centering
\includegraphics[width=0.75\textwidth]{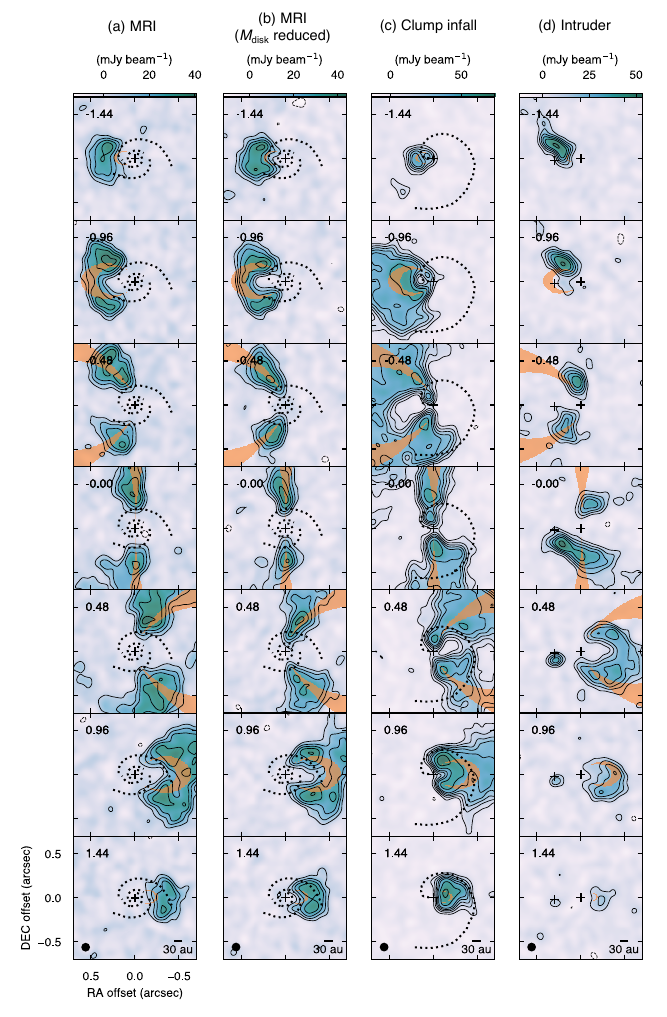}
\caption{Velocity channel maps of the $\CeighteenO$ $J=3$--2 emission after beam convolution (green) overlaid with Keplerian iso-velocity curves (orange) expected for the protostellar mass. The channel maps are presented in steps of $0.48~\kmps$, while the original velocity resolution of the synthetic data is $0.16~\kmps$. Contour levels are 3, 6, 9, 12, 15, and 20 $\times \sigma$, where $\sigma = 2.5~\mjypbm$. Dotted curves denote the locations of the spiral arms, and crosses indicate stellar positions. Filled circles at the bottom left corners show the beam size of $0\farcs1$.}
\label{fig:channel_all_withbeam}
\end{figure*}

\subsubsection{Beam-convolved maps}\label{subsec:beamconv}

To assess how the kinematic features discussed above would appear in actual observations, we present synthetic images convolved by a Gaussian beam with random noise added. In particular, we focus on identifying key observational differences among the three models that could help distinguish them.


Figure \ref{fig:mom1_beam} presents the mean-velocity and residual velocity maps of the \ce{C^18O} emission after beam convolution. The mean-velocity maps were computed by integrating emission detected at $4\sigma$ or more. In the MRI model, no local kinematic features indicative of GI, such as repeated blueshifted and redshifted residuals along the disk minor axis, are visible. The emission is absent near the disk center, as found in the velocity channel maps without beam convolution and noise (Figure \ref{fig:channel_all}a). The central hole appears slightly larger in the beam-convolved map, likely due to weak line emission being obscured by noise in regions dominated by the optically thick dust continuum.

To examine the case where the dust continuum is less optically thick and the \ce{C^18O} emission is detected at inner radii,  we artificially reduced the disk mass in the MRI model by a factor of three and ran the radiative transfer calculation. The resultant disk mass of $0.12~\Msun$ still satisfies the typical condition of $M_\mathrm{disk} / M_\ast \gtrsim 0.1$ for the gravitationally unstable disk \citep{Kratter2016}. The mean-velocity and residual velocity maps for the modified MRI model are presented in the second column of Figure \ref{fig:mom1_beam}. Although the modified model exhibits stronger emission in the inner disk, GI features remain difficult to identify, indicating that they are completely smoothed out at the assumed distance and angular resolution. We note that wavy, mosaic-like patterns enhanced in the residual maps of the original and modified MRI models are artificial features caused by the limited number of channels that are used to compute the mean-velocity map under the current velocity resolution and signal-to-noise ratio (S/N), but they do not affect the detection of GI wiggles. We confirmed that the overall velocity structures remain the same at higher velocity resolution and S/N, and the GI features are undetectable even when the wavy patterns are suppressed.

In contrast, the clump-infall model retains clear signatures of expanding gas motion along the spiral in the residual map. This spiral-shaped expansion feature provides a robust diagnostic for distinguishing the clump-infall scenario from the other two models. 
Similarly, the intruder model continues to show strong velocity perturbations, as seen in the residual velocity map without beam convolution.

The residual velocity maps discussed above were computed using projected Keplerian velocity fields with the true protostellar mass adopted from the hydrodynamical simulations, and the inclination angle fixed in the radiative transfer calculations. In real observations, however, these parameters must be inferred from data and are subject to uncertainties. To assess how uncertainties in the protostellar mass and inclination angle affect the residual velocity structure, we recalculated residual velocity maps using projected Keplerian velocity fields with intentionally offset parameter values. Figure \ref{fig:mom1_mserror} in Appendix \ref{app:fig_suppl} presents residual velocity maps calculated with protostellar masses that are $20\%$ smaller or larger than the true value, which represents a conservative estimate of the typical uncertainty in protostellar mass measurements based on disk rotation \citep[e.g.,][]{Simon:2000aa, Pietu:2007aa, Guilloteau:2014aa, Sheehan:2019aa, Aso:2020aa}. The residual velocities of $\gtrsim 0.5~\kmps$ associated with the expanding spiral in the clump-infall model and with accretion by the intruder in the intruder model remain clearly identifiable even with a 20\% error in the protostellar mass. Similarly, adopting inclination angles offset by $\rtsim3^\circ$, which is a typical uncertainty in inclination angle estimates based on modeling spatially resolved Keplerian velocity patterns \citep[e.g.,][]{Simon:2000aa, Pietu:2007aa, Guilloteau:2014aa, Sheehan:2019aa}, also introduces only a minor error in residual velocities compared to the velocity perturbations induced by the spiral or the intruder. Hence, these parameter deviations do not significantly impact the detectability of the key kinematic signatures expected for either the clump-infall or intruder scenario.

To evaluate the dependence of these kinematic features on viewing angles, i.e., the azimuthal location of the spirals or intruder and the inclination angle, we also produced synthetic line images adopting different azimuthal and inclination angles. Residual velocity maps for all models with rotational angles of $90^\circ, 180^\circ$, and $270^\circ$ are presented in Figure \ref{fig:mom1_mri_rotated}--\ref{fig:mom1_intruder_rotated} in Appendix \ref{app:fig_suppl}. These maps confirm that the key kinematic features discussed above remain identifiable regardless of azimuthal viewing angles when the disk is viewed nearly face-on ($i=30^\circ$). Residual velocity maps for models with inclination angles of $45^\circ$ and $60^\circ$ are shown in Figure \ref{fig:mom1_more_inclined} in the Appendix. When the inclination angle increases to $60^\circ$, additional residual velocity structures that are apart from the key features highlighted above become apparent, particularly in the MRI and clump-infall models, where the disks are surrounded by infalling envelopes. These additional residuals arise because the hydrodynamical models have a finite thickness, which increases with distance due to flaring, leading to multiple velocity components along the line of sight at higher inclination angles. Therefore, disks having small but nonzero inclination angles are most suitable for investigating disk kinematics, as also indicated by previous detailed studies of gas kinematics in Class \II~disks \citep[][]{Teague:2025aa}.

Figure \ref{fig:channel_all_withbeam} presents velocity channel maps of the \ce{C^18O} emission after beam convolution and inclusion of noise. Expected Keplerian velocity patterns calculated from the protostellar masses listed in Table \ref{tbl:initial} and an inclination angle of $30^\circ$ are also  overlaid. The \ce{C^18O} emission in both the original and modified MRI models generally follows Keplerian rotation (Figure \ref{fig:channel_all_withbeam}a and b). The GI wiggles found in Figure \ref{fig:channel_all}a are smoothed out and are no longer clearly visible after beam convolution. 
The clump-infall model exhibits somewhat broader channel maps. The intensity becomes weaker at the position of the inner spiral arm due to its lower temperature, resulting in a neck-like structure. 
Overall, both the MRI and clump-infall models exhibit Keplerian rotation patterns, making them difficult to distinguish based solely on the channel maps. In contrast, the intruder model shows a clear departure from Keplerian motion even after beam convolution.

When comparing the velocity channel maps of the three models, the intruder model possesses some characteristic features that can be used to distinguish it from the MRI and clump-infall models. In particular, channels with equal absolute velocity values show pronounced asymmetry when compared to each other, and the channel centered at zero velocity is strongly asymmetric with respect to the disk's minor axis. These effects are introduced by the intruder and are much less expressed in the other two models.

In summary, the clump-infall and intruder models exhibit unique kinematic features that can be used to differentiate them from the others at a spatial resolution of $30~\au$. The clump-infall model is characterized by a feature of spiral-shaped expansion in the residual velocity map, which arises due to the exchange of the angular momentum between the infalling clump and the spiral. The intruder model shows a highly asymmetric velocity structure with respect to the systemic velocity of the primary disk in the channel maps. These features are observable in nearby FUors ($d=300~\pc$) with ALMA, as the assumed angular and velocity resolutions are achievable with reasonably good sensitivity even for less abundant molecular tracers such as \ce{C^18O}. In contrast, the MRI model lacks prominent local kinematic features in both the residual or channel maps, because GI-induced structures are smoothed out by the observing beam. Hence, the absence of the above two kinematic features suggests that a combination of MRI and GI may operate. These distinct kinematic signatures offer promising diagnostics for identifying the physical mechanisms driving FU Orionis-type outbursts.


\section{Conclusions} \label{sec:conclusion}

We investigated observational signatures of the gas morphology and kinematics associated with three FU Orionis-type accretion outburst models---the combination of the GI and MRI, clump infall, and stellar intrusion---using synthetic observations of hydrodynamical models in the \ce{C^18O} $J=3\mbox{--}2$ line emission at a distance of $300~\pc$. Our main conclusions are summarized below:

\begin{itemize}
    \item The integrated-intensity maps of the synthetic \ce{C^18O} $J=3\mbox{--}2$ show that line intensity does not reliably trace high-density spirals formed by GI or infalling clumps in outburst-triggering disks because of their peculiar gas temperature distributions. However, the clump-infall and intruder models exhibit asymmetric intensity distributions due to asymmetric temperature and density structures of the disks, which may help distinguish them from the MRI model when combined with kinematic diagnostics.
    \item Distinct kinematic features are evident in the residual velocity maps, where projected Keplerian velocity fields are subtracted from the mean-velocity maps. In the MRI model, gas gravitationally perturbed by GI-induced spirals produces repeated blueshifted and redshifted residuals of $\rtsim0.1~\kmps$ on scales of several au along the disk minor axis. The clump-infall model shows a feature of spiral-shaped expansion with residuals of $\rtsim0.5~\kmps$ across several tens au, which is caused by angular momentum exchange between the infalling clump and surrounding gas. The intruder model exhibits strong residual velocities of $\gtrsim 1~\kmps$ around the intruder, which are attributed to accretion by the intruder.
    \item In the velocity channel maps, each model exhibits following characteristic features. The MRI model shows kinks, so-called GI wiggles, on scales of several au, caused by local velocity perturbations around GI-induced spirals. The clump-infall model exhibits similar but stronger kinks near the inner spiral arm formed by an infalling clump. The intruder model shows highly perturbed velocity structures on the intruder side of the disk, while the opposite side retains a relatively undisturbed Keplerian pattern.
    \item After applying convolution of a Gaussian beam with FWHM of $0\farcs1$ and adding random noise of $2.5~\mjypbm$, the following kinematic characteristics are found to be useful for distinguishing the three models. GI wiggles in the MRI model are completely smoothed out, leaving no prominent local kinematic features both in the residual and channel maps. In contrast, the clump-infall model retains a clear sign of spiral-shaped expansion in the residual velocity map. The intruder model shows asymmetric velocity structures with respect to the systemic velocity of the primary disk in the velocity channel maps. These kinematic characteristics observable with ALMA  at a spatial resolution of $30~\au$ complement morphological information from continuum observations and offer promising diagnostics for identifying the physical mechanisms behind FU Orionis-type outbursts.
\end{itemize}

\begin{acknowledgments}
We thank the anonymous referee for providing an insightful review that helped to improve the paper. J.S. acknowledge support from the KU-DREAM program of Kagoshima University. E.I.V. and A.S. acknowledge support from the Ministry of Science and Higher Education of the Russian Federation (State assignment in the field of scientific activity 2023, GZ0110/23-10-IF). Simulations were performed on the Vienna Scientific Cluster (VSC).
\end{acknowledgments}

\appendix

\section{Temperature inversion during the burst}
\label{App:temp-inv}

\begin{figure}
\begin{centering}
\includegraphics[width=\linewidth]{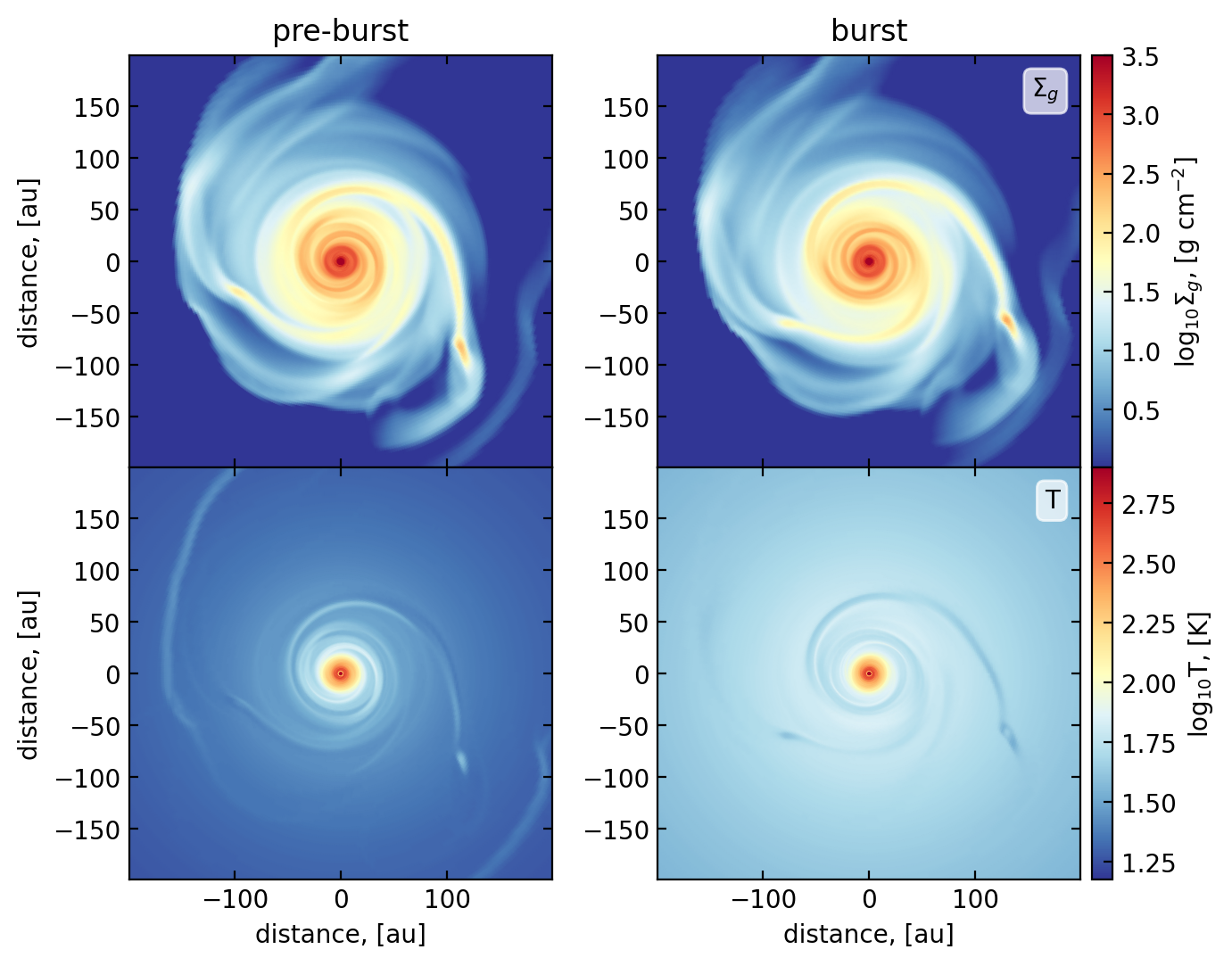} 
\par \end{centering}
\caption{Gas surface density (top row) and temperature (bottom row) just before the burst (left column) and during the burst (right column), namely, 30 yr after the burst onset in the MRI model.} 
\label{fig:tempr-inv}
\end{figure}

Figure~\ref{fig:tempr-inv} compares the spatial distributions of the gas surface density and temperature in the pre-burst and burst phases of the MRI model. The gas surface density changes only slightly, whereas the temperature notably increases throughout most of the disk due to increased stellar irradiation during the burst. The temperature inversion in the vicinity of the spiral arms is clearly visible. The spiral arms are systematically warmer than the interspiral regions before the burst due to compressional heating caused by the spiral density wave,  whereas during the burst the opposite trend is observed. This temperature inversion effect is therefore characteristic of the burst in its early development stage ($\le 30$~yr). For much longer bursts, the temperature inversion may diminish and this deserves further investigation in future works.

\section{Geometry-thin MRI Model} \label{app:mri_thin}

\begin{figure}[htbp]
\centering
\includegraphics[width=0.5\textwidth]{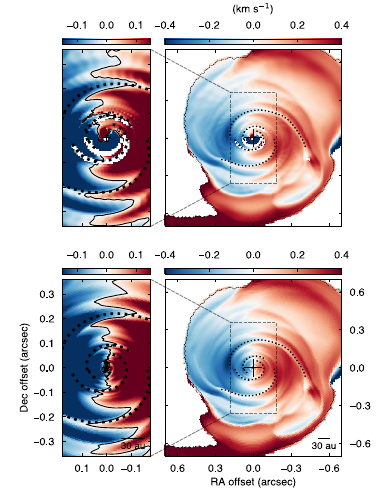}
\caption{Same as the third and fourth rows of Figure \ref{fig:mom1_all} but for a geometry thin MRI model.}
\label{fig:mom1_mri_thin}
\end{figure}

To examine whether less clear GI features in the residual map of the \ce{C^18O} emission of the MRI model is due to the disk's vertical thickness, we artificially reduced the disk scale height by a factor of $0.1$ and ran a radiative transfer calculation. The residual map of the \ce{C^18O} $J=3\mbox{--}2$ emission of the geometry-thin MRI model is presented in Figure \ref{fig:mom1_mri_thin}. The geometry-thin model retains the repeated redshifted and blueshifted patterns along the disk minor axis in the residual velocity map. The velocity structures are almost identical to the residual map of the raw data except for those at very inner radii, where the absorption by the dust continuum emission is strong. These results suggest that the less clear GI features in the \ce{C^18O} emission of the original MRI model are attributed to the smoothing effects by the disk's vertical thickness.

\section{Supplemental Figures} \label{app:fig_suppl}
\begin{figure*}[tbhp]
\centering
\includegraphics[width=0.82\textwidth]{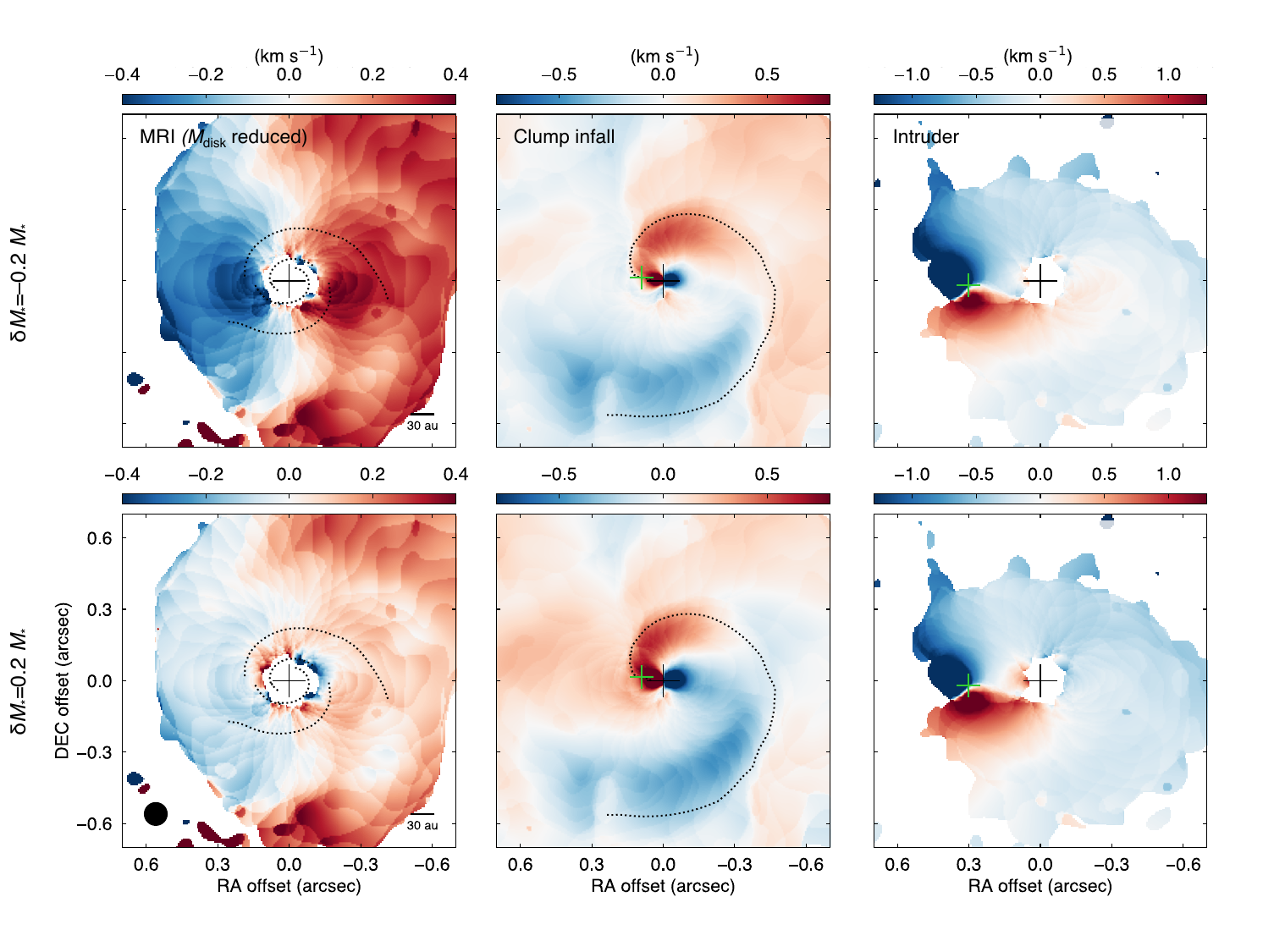}
\caption{Residual velocity maps of the beam-convolved \ce{C^18O} $J=3\mbox{--}2$ emission, calculated using Keplerian velocity fields with stellar masses smaller (top) and larger (bottom) than the true values by $20\%$. The dotted curves indicate the locations of the high-density spiral arms. The filled circle in the bottom left panel shows the beam size of $0\farcs1$. Black crosses indicate the protostellar positions, and green crosses denote positions of the inner clump or intruder.}
\label{fig:mom1_mserror}
\end{figure*}

\begin{figure*}[tbhp]
\centering
\includegraphics[width=0.82\textwidth]{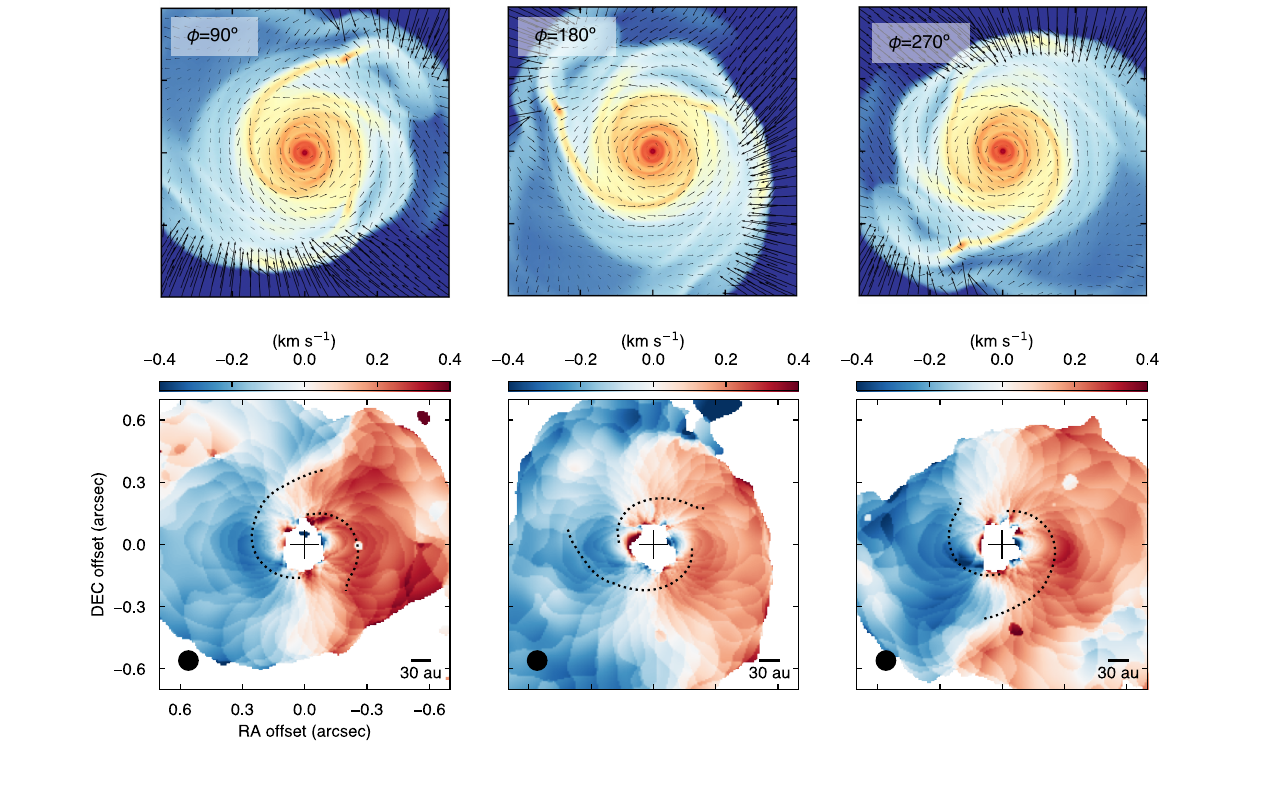}
\caption{The model surface density distributions (top) and residual velocity maps (bottom) for the modified MRI model rotated by an angle of $\phi$ in the counter clockwise direction. We note that rotation by the inclination angle is not included in the surface density plots. The dotted curves and filled circles in the residual maps denote the locations of the high-density spiral arms and the beam size of $0\farcs1$, respectively.}
\label{fig:mom1_mri_rotated}
\end{figure*}

\begin{figure*}[tbhp]
\centering
\includegraphics[width=0.82\textwidth]{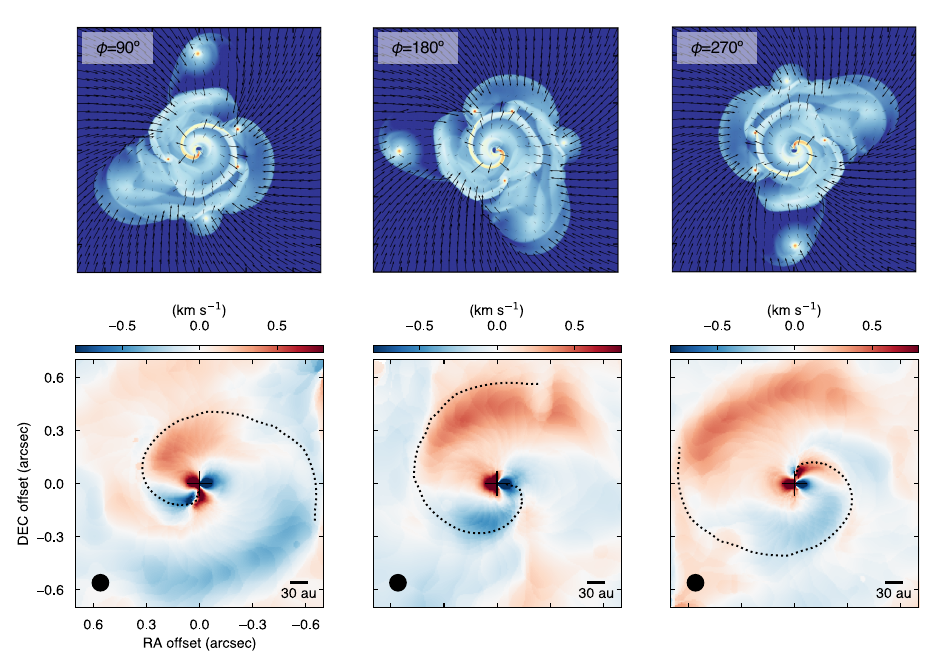}
\caption{Same as Figure \ref{fig:mom1_mri_rotated} but for the clump-infall model.}
\label{fig:mom1_clump_rotated}
\end{figure*}

\begin{figure*}[tbhp]
\centering
\includegraphics[width=0.82\textwidth]{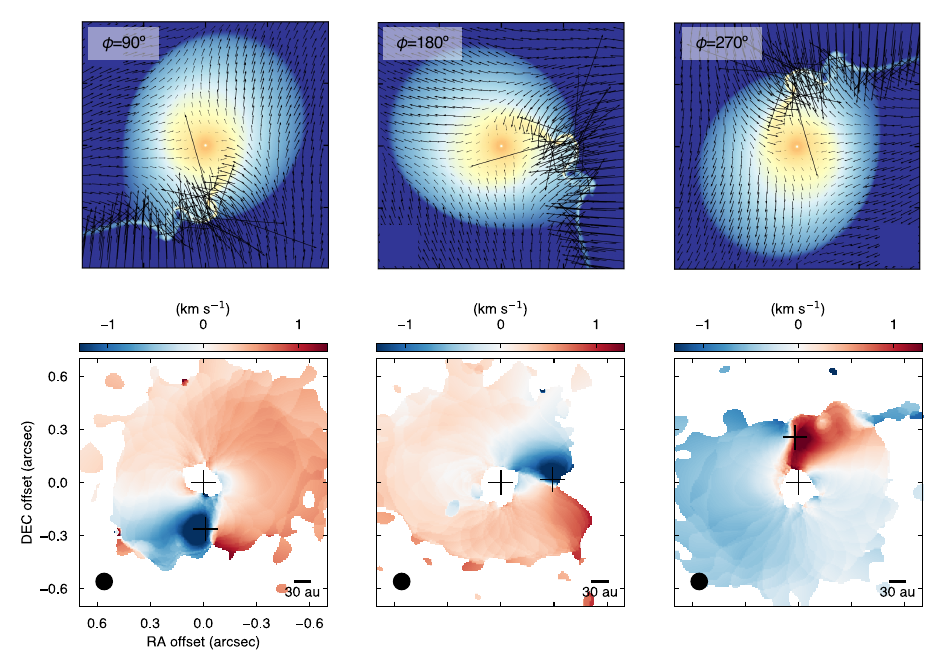}
\caption{Same as Figure \ref{fig:mom1_mri_rotated} but for the intruder model.}
\label{fig:mom1_intruder_rotated}
\end{figure*}

\begin{figure*}[tbhp]
\centering
\includegraphics[width=0.82\textwidth]{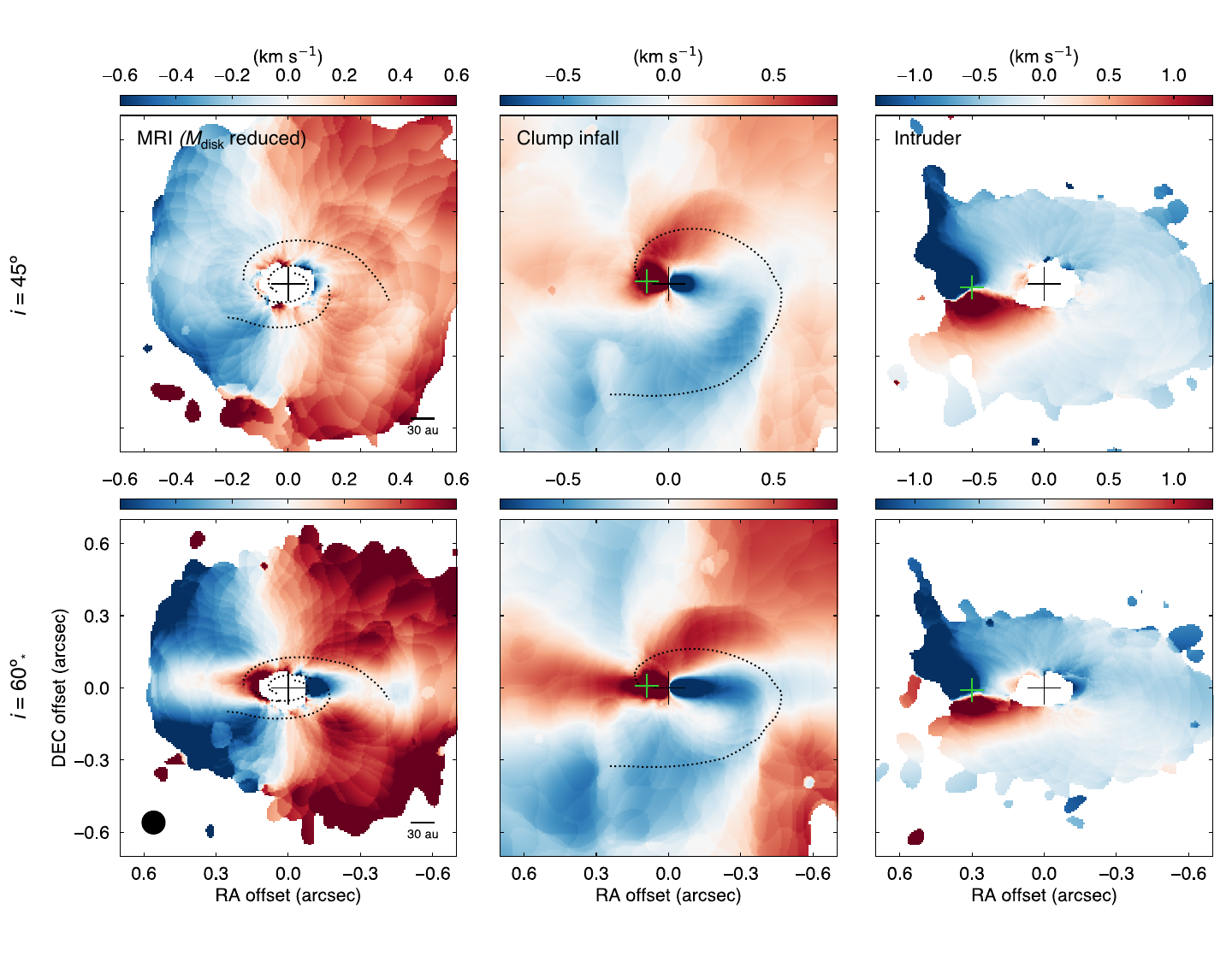}
\caption{Residual velocity maps of the beam-convolved \ce{C^18O} $J=3\mbox{--}2$ emission of the modified MRI, clump-infall and intruder models with inclination angles of $45^\circ$ (top) and $60^\circ$ (bottom). The dotted curves indicate the locations of the high-density spiral arms. The filled circle in the bottom left panel shows the beam size of $0\farcs1$. Black crosses indicate the protostellar positions, and green crosses denote positions of the inner clump or intruder.}
\label{fig:mom1_more_inclined}
\end{figure*}

Figure \ref{fig:mom1_mserror} presents residual velocity maps computed using projected Keplerian velocity fields with protostellar masses offset from the true values by $20\%$. Although not shown, adopting an inclination error of $\pm3^\circ$ produces nearly identical results to those in Figure \ref{fig:mom1_mserror}. Residual velocity maps for the three models with different azimuthal angles and inclination angles are presented in Figure \ref{fig:mom1_mri_rotated}--\ref{fig:mom1_more_inclined}. In all Figures \ref{fig:mom1_mserror}--\ref{fig:mom1_more_inclined}, we used the modified MRI model, where the disk mass is reduced by a factor of three, instead of the original MRI model to suppress the optical depth of the dust continuum near the center.

\bibliography{reference, reference-2}{}

\begin{thebibliography}{}
\expandafter\ifx\csname natexlab\endcsname\relax\def\natexlab#1{#1}\fi
\providecommand{\url}[1]{\href{#1}{#1}}
\providecommand{\dodoi}[1]{doi:~\href{http://doi.org/#1}{\nolinkurl{#1}}}
\providecommand{\doeprint}[1]{\href{http://ascl.net/#1}{\nolinkurl{http://ascl.net/#1}}}
\providecommand{\doarXiv}[1]{\href{https://arxiv.org/abs/#1}{\nolinkurl{https://arxiv.org/abs/#1}}}

\bibitem[{P.~J. {Armitage} {et~al.}(2001){Armitage}, {Livio}, \& {Pringle}}]{Armitage2001}
{Armitage}, P.~J., {Livio}, M., \& {Pringle}, J.~E. 2001, \bibinfo{title}{{Episodic accretion in magnetically layered protoplanetary discs},} \mnras, 324, 705, \dodoi{10.1046/j.1365-8711.2001.04356.x}

\bibitem[{Y. {Aso} \& M.~N. {Machida}(2020){Aso} \& {Machida}}]{Aso:2020aa}
{Aso}, Y., \& {Machida}, M.~N. 2020, \bibinfo{title}{{Testing Disk Identification Methods through Numerical Simulations of Protostellar Evolution},} \apj, 905, 174

\bibitem[{M. {Audard} {et~al.}(2014){Audard}, {{\'A}brah{\'a}m}, {Dunham}, {Green}, {Grosso}, {Hamaguchi}, {Kastner}, {K{\'o}sp{\'a}l}, {Lodato}, {Romanova}, {Skinner}, {Vorobyov}, \& {Zhu}}]{Audard:2014aa}
{Audard}, M., {{\'A}brah{\'a}m}, P., {Dunham}, M.~M., {et~al.} 2014, in Protostars and Planets VI, ed. H.~{Beuther}, R.~S. {Klessen}, C.~P. {Dullemond}, \& T.~{Henning}, 387--410

\bibitem[{J. {Bae} {et~al.}(2014){Bae}, {Hartmann}, {Zhu}, \& {Nelson}}]{Bae2014}
{Bae}, J., {Hartmann}, L., {Zhu}, Z., \& {Nelson}, R.~P. 2014, \bibinfo{title}{{Accretion Outbursts in Self-gravitating Protoplanetary Disks},} \apj, 795, 61, \dodoi{10.1088/0004-637X/795/1/61}

\bibitem[{T. {Birnstiel} {et~al.}(2018){Birnstiel}, {Dullemond}, {Zhu}, {Andrews}, {Bai}, {Wilner}, {Carpenter}, {Huang}, {Isella}, {Benisty}, {P{\'e}rez}, \& {Zhang}}]{Birnstiel:2018aa}
{Birnstiel}, T., {Dullemond}, C.~P., {Zhu}, Z., {et~al.} 2018, \bibinfo{title}{{The Disk Substructures at High Angular Resolution Project (DSHARP). V. Interpreting ALMA Maps of Protoplanetary Disks in Terms of a Dust Model},} \apjl, 869, L45

\bibitem[{E.~M.~A. Borchert {et~al.}(2021)Borchert, Price, Pinte, \& Cuello}]{Borchert2021a}
Borchert, E. M.~A., Price, D.~J., Pinte, C., \& Cuello, N. 2021, \bibinfo{title}{On the rise times in {FU} {Orionis} events,} Monthly Notices of the Royal Astronomical Society: Letters, 510, L37, \dodoi{10.1093/mnrasl/slab123}

\bibitem[{E.~M.~A. {Borchert} {et~al.}(2022){Borchert}, {Price}, {Pinte}, \& {Cuello}}]{Borchert2022}
{Borchert}, E. M.~A., {Price}, D.~J., {Pinte}, C., \& {Cuello}, N. 2022, \bibinfo{title}{{Sustained FU Orionis-type outbursts from colliding discs in stellar flybys},} \mnras, 517, 4436, \dodoi{10.1093/mnras/stac2872}

\bibitem[{P. {Colella} \& P.~R. {Woodward}(1984){Colella} \& {Woodward}}]{Colella1984}
{Colella}, P., \& {Woodward}, P.~R. 1984, \bibinfo{title}{{The Piecewise Parabolic Method (PPM) for Gas-Dynamical Simulations},} Journal of Computational Physics, 54, 174, \dodoi{10.1016/0021-9991(84)90143-8}

\bibitem[{M.~S. {Connelley} \& B. {Reipurth}(2018){Connelley} \& {Reipurth}}]{Connelley:2018aa}
{Connelley}, M.~S., \& {Reipurth}, B. 2018, \bibinfo{title}{{A Near-infrared Spectroscopic Survey of FU Orionis Objects},} \apj, 861, 145

\bibitem[{N. {Cuello} {et~al.}(2023){Cuello}, {M{\'e}nard}, \& {Price}}]{Cuello2023}
{Cuello}, N., {M{\'e}nard}, F., \& {Price}, D.~J. 2023, \bibinfo{title}{{Close encounters: How stellar flybys shape planet-forming discs},} European Physical Journal Plus, 138, 11, \dodoi{10.1140/epjp/s13360-022-03602-w}

\bibitem[{T.~V. {Demidova} \& V.~P. {Grinin}(2023){Demidova} \& {Grinin}}]{Demidova2023}
{Demidova}, T.~V., \& {Grinin}, V.~P. 2023, \bibinfo{title}{{Three-dimensional SPH Simulations of FU Orionis Star Flares in the Clumpy Accretion Model},} \apj, 953, 38, \dodoi{10.3847/1538-4357/acdf5f}

\bibitem[{R. {Dong} {et~al.}(2022){Dong}, {Liu}, {Cuello}, {Pinte}, {{\'A}brah{\'a}m}, {Vorobyov}, {Hashimoto}, {K{\'o}sp{\'a}l}, {Chiang}, {Takami}, {Chen}, {Dunham}, {Fukagawa}, {Green}, {Hasegawa}, {Henning}, {Pavlyuchenkov}, {Pyo}, \& {Tamura}}]{Dong:2022aa}
{Dong}, R., {Liu}, H.~B., {Cuello}, N., {et~al.} 2022, \bibinfo{title}{{A likely flyby of binary protostar Z CMa caught in action},} Nature Astronomy, 6, 331

\bibitem[{C.~P. {Dullemond} {et~al.}(2019){Dullemond}, {K{\"u}ffmeier}, {Goicovic}, {Fukagawa}, {Oehl}, \& {Kramer}}]{Dullemond2019}
{Dullemond}, C.~P., {K{\"u}ffmeier}, M., {Goicovic}, F., {et~al.} 2019, \bibinfo{title}{{Cloudlet capture by transitional disk and FU Orionis stars},} \aap, 628, A20, \dodoi{10.1051/0004-6361/201832632}

\bibitem[{M.~M. {Dunham} \& E.~I. {Vorobyov}(2012){Dunham} \& {Vorobyov}}]{Dunham2012}
{Dunham}, M.~M., \& {Vorobyov}, E.~I. 2012, \bibinfo{title}{{Resolving the Luminosity Problem in Low-mass Star Formation},} \apj, 747, 52, \dodoi{10.1088/0004-637X/747/1/52}

\bibitem[{N.~J. {Evans} {et~al.}(2009){Evans}, {Dunham}, {J{\o}rgensen}, {Enoch}, {Mer{\'{\i}}n}, {van Dishoeck}, {Alcal{\'a}}, {Myers}, {Stapelfeldt}, {Huard}, {Allen}, {Harvey}, {van Kempen}, {Blake}, {Koerner}, {Mundy}, {Padgett}, \& {Sargent}}]{Evans:2009aa}
{Evans}, II, N.~J., {Dunham}, M.~M., {J{\o}rgensen}, J.~K., {et~al.} 2009, \bibinfo{title}{{The Spitzer c2d Legacy Results: Star-Formation Rates and Efficiencies; Evolution and Lifetimes},} \apjs, 181, 321

\bibitem[{W.~J. {Fischer} {et~al.}(2023){Fischer}, {Hillenbrand}, {Herczeg}, {Johnstone}, {Kospal}, \& {Dunham}}]{Fischer:2023aa}
{Fischer}, W.~J., {Hillenbrand}, L.~A., {Herczeg}, G.~J., {et~al.} 2023, in Astronomical Society of the Pacific Conference Series, Vol. 534, Protostars and Planets VII, ed. S.~{Inutsuka}, Y.~{Aikawa}, T.~{Muto}, K.~{Tomida}, \& M.~{Tamura}, 355

\bibitem[{D. {Forgan} \& K. {Rice}(2010){Forgan} \& {Rice}}]{Forgan2010}
{Forgan}, D., \& {Rice}, K. 2010, \bibinfo{title}{{Stellar encounters in the context of outburst phenomena},} \mnras, 402, 1349, \dodoi{10.1111/j.1365-2966.2009.15974.x}

\bibitem[{M.~A. {Frerking} {et~al.}(1982){Frerking}, {Langer}, \& {Wilson}}]{Frerking:1982aa}
{Frerking}, M.~A., {Langer}, W.~D., \& {Wilson}, R.~W. 1982, \bibinfo{title}{{The relationship between carbon monoxide abundance and visual extinction in interstellar clouds},} \apj, 262, 590

\bibitem[{C.~F. {Gammie}(1996){Gammie}}]{Gammie1996}
{Gammie}, C.~F. 1996, \bibinfo{title}{{Layered Accretion in T Tauri Disks},} \apj, 457, 355, \dodoi{10.1086/176735}

\bibitem[{S. {Guilloteau} {et~al.}(2014){Guilloteau}, {Simon}, {Pi{\'e}tu}, {Di Folco}, {Dutrey}, {Prato}, \& {Chapillon}}]{Guilloteau:2014aa}
{Guilloteau}, S., {Simon}, M., {Pi{\'e}tu}, V., {et~al.} 2014, \bibinfo{title}{{The masses of young stars: CN as a probe of dynamical masses},} \aap, 567, A117

\bibitem[{C. {Hall} {et~al.}(2019){Hall}, {Dong}, {Rice}, {Harries}, {Najita}, {Alexander}, \& {Brittain}}]{Hall:2019aa}
{Hall}, C., {Dong}, R., {Rice}, K., {et~al.} 2019, \bibinfo{title}{{The Temporal Requirements of Directly Observing Self-gravitating Spiral Waves in Protoplanetary Disks with ALMA},} \apj, 871, 228

\bibitem[{C. {Hall} {et~al.}(2020){Hall}, {Dong}, {Teague}, {Terry}, {Pinte}, {Paneque-Carre{\~n}o}, {Veronesi}, {Alexander}, \& {Lodato}}]{Hall:2020aa}
{Hall}, C., {Dong}, R., {Teague}, R., {et~al.} 2020, \bibinfo{title}{{Predicting the Kinematic Evidence of Gravitational Instability},} \apj, 904, 148

\bibitem[{A.~F. {Izquierdo} {et~al.}(2025){Izquierdo}, {Stadler}, {Galloway-Sprietsma}, {Benisty}, {Pinte}, {Bae}, {Teague}, {Facchini}, {W{\"o}lfer}, {Longarini}, {Curone}, {Andrews}, {Barraza-Alfaro}, {Cataldi}, {Cuello}, {Czekala}, {Fasano}, {Flock}, {Fukagawa}, {Garg}, {Hall}, {Hammond}, {Hilder}, {Huang}, {Ilee}, {Isella}, {Kanagawa}, {Lesur}, {Lodato}, {Loomis}, {Orihara}, {Price}, {Rosotti}, {Testi}, {Yen}, {Wafflard-Fernandez}, {Wilner}, {Winter}, {Yoshida}, \& {Zawadzki}}]{Izquierdo:2025aa}
{Izquierdo}, A.~F., {Stadler}, J., {Galloway-Sprietsma}, M., {et~al.} 2025, \bibinfo{title}{{exoALMA. III. Line-intensity Modeling and System Property Extraction from Protoplanetary Disks},} \apjl, 984, L8

\bibitem[{S.~J. {Kenyon} {et~al.}(1990){Kenyon}, {Hartmann}, {Strom}, \& {Strom}}]{Kenyon1990}
{Kenyon}, S.~J., {Hartmann}, L.~W., {Strom}, K.~M., \& {Strom}, S.~E. 1990, \bibinfo{title}{{An IRAS Survey of the Taurus-Auriga Molecular Cloud},} \aj, 99, 869, \dodoi{10.1086/115380}

\bibitem[{K. {Kratter} \& G. {Lodato}(2016){Kratter} \& {Lodato}}]{Kratter2016}
{Kratter}, K., \& {Lodato}, G. 2016, \bibinfo{title}{{Gravitational Instabilities in Circumstellar Disks},} \araa, 54, 271, \dodoi{10.1146/annurev-astro-081915-023307}

\bibitem[{S.~I. {Laznevoi} {et~al.}(2025){Laznevoi}, {Akimkin}, {Pavlyuchenkov}, {Il'in}, {K{\'o}sp{\'a}l}, \& {{\'A}brah{\'a}m}}]{Laznevoi2025}
{Laznevoi}, S.~I., {Akimkin}, V.~V., {Pavlyuchenkov}, Y.~N., {et~al.} 2025, \bibinfo{title}{{Time-dependent response of protoplanetary disk temperature to an FU Ori-type luminosity outburst},} \aap, 700, L24, \dodoi{10.1051/0004-6361/202554962}

\bibitem[{H.~B. {Liu} {et~al.}(2016){Liu}, {Takami}, {Kudo}, {Hashimoto}, {Dong}, {Vorobyov}, {Pyo}, {Fukagawa}, {Tamura}, {Henning}, {Dunham}, {Karr}, {Kusakabe}, \& {Tsuribe}}]{Liu:2016aa}
{Liu}, H.~B., {Takami}, M., {Kudo}, T., {et~al.} 2016, \bibinfo{title}{{Circumstellar disks of the most vigorously accreting young stars},} Science Advances, 2, e1500875

\bibitem[{H.~B. {Liu} {et~al.}(2017){Liu}, {Vorobyov}, {Dong}, {Dunham}, {Takami}, {Galv{\'a}n-Madrid}, {Hashimoto}, {K{\'o}sp{\'a}l}, {Henning}, {Tamura}, {Rodr{\'\i}guez}, {Hirano}, {Hasegawa}, {Fukagawa}, {Carrasco-Gonzalez}, \& {Tazzari}}]{Liu2017}
{Liu}, H.~B., {Vorobyov}, E.~I., {Dong}, R., {et~al.} 2017, \bibinfo{title}{{A concordant scenario to explain FU Orionis from deep centimeter and millimeter interferometric observations},} \aap, 602, A19, \dodoi{10.1051/0004-6361/201630263}

\bibitem[{M.~N. {Machida} {et~al.}(2011){Machida}, {Inutsuka}, \& {Matsumoto}}]{Machida2011}
{Machida}, M.~N., {Inutsuka}, S.-i., \& {Matsumoto}, T. 2011, \bibinfo{title}{{Recurrent Planet Formation and Intermittent Protostellar Outflows Induced by Episodic Mass Accretion},} \apj, 729, 42, \dodoi{10.1088/0004-637X/729/1/42}

\bibitem[{D.~M.-A. {Meyer} {et~al.}(2017){Meyer}, {Vorobyov}, {Kuiper}, \& {Kley}}]{2017MeyerVorobyov}
{Meyer}, D.~M.-A., {Vorobyov}, E.~I., {Kuiper}, R., \& {Kley}, W. 2017, \bibinfo{title}{{On the existence of accretion-driven bursts in massive star formation},} \mnras, 464, L90, \dodoi{10.1093/mnrasl/slw187}

\bibitem[{S. {Nayakshin} \& G. {Lodato}(2012){Nayakshin} \& {Lodato}}]{Nayakshin:2012aa}
{Nayakshin}, S., \& {Lodato}, G. 2012, \bibinfo{title}{{Fu Ori outbursts and the planet-disc mass exchange},} \mnras, 426, 70

\bibitem[{S. {Nayakshin} {et~al.}(2023){Nayakshin}, {Owen}, \& {Elbakyan}}]{Nayakshin2023}
{Nayakshin}, S., {Owen}, J.~E., \& {Elbakyan}, V. 2023, \bibinfo{title}{{Extreme evaporation of planets in hot thermally unstable protoplanetary discs: the case of FU Ori},} \mnras, 523, 385, \dodoi{10.1093/mnras/stad1392}

\bibitem[{N. {Ohashi} {et~al.}(2023){Ohashi}, {Tobin}, {J{\o}rgensen}, {Takakuwa}, {Sheehan}, {Aikawa}, {Li}, {Looney}, {Williams}, {Aso}, {Sharma}, {Choi}, {Yamato}, {Lee}, {Tomida}, {Yen}, {Encalada}, {Flores}, {Gavino}, {Kido}, {Han}, {Lin}, {Narayanan}, {Phuong}, {Santamar{\'\i}a-Miranda}, {Thieme}, {van't Hoff}, {de Gregorio-Monsalvo}, {Koch}, {Kwon}, {Lai}, {Lee}, {Plunkett}, {Saigo}, {Hirano}, {Lam}, \& {Mori}}]{Ohashi:2023aa}
{Ohashi}, N., {Tobin}, J.~J., {J{\o}rgensen}, J.~K., {et~al.} 2023, \bibinfo{title}{{Early Planet Formation in Embedded Disks (eDisk). I. Overview of the Program and First Results},} \apj, 951, 8

\bibitem[{V. {Pi{\'e}tu} {et~al.}(2007){Pi{\'e}tu}, {Dutrey}, \& {Guilloteau}}]{Pietu:2007aa}
{Pi{\'e}tu}, V., {Dutrey}, A., \& {Guilloteau}, S. 2007, \bibinfo{title}{{Probing the structure of protoplanetary disks: a comparative study of DM Tau, LkCa 15, and MWC 480},} \aap, 467, 163

\bibitem[{C. {Pinte} {et~al.}(2018){Pinte}, {Price}, {M{\'e}nard}, {Duch{\^e}ne}, {Dent}, {Hill}, {de Gregorio-Monsalvo}, {Hales}, \& {Mentiplay}}]{Pinte:2018aa}
{Pinte}, C., {Price}, D.~J., {M{\'e}nard}, F., {et~al.} 2018, \bibinfo{title}{{Kinematic Evidence for an Embedded Protoplanet in a Circumstellar Disk},} \apjl, 860, L13

\bibitem[{P.~D. {Sheehan} {et~al.}(2019){Sheehan}, {Wu}, {Eisner}, \& {Tobin}}]{Sheehan:2019aa}
{Sheehan}, P.~D., {Wu}, Y.-L., {Eisner}, J.~A., \& {Tobin}, J.~J. 2019, \bibinfo{title}{{High-precision Dynamical Masses of Pre-main-sequence Stars with ALMA and Gaia},} \apj, 874, 136

\bibitem[{M. {Simon} {et~al.}(2000){Simon}, {Dutrey}, \& {Guilloteau}}]{Simon:2000aa}
{Simon}, M., {Dutrey}, A., \& {Guilloteau}, S. 2000, \bibinfo{title}{{Dynamical Masses of T Tauri Stars and Calibration of Pre-Main-Sequence Evolution},} \apj, 545, 1034

\bibitem[{A.~M. {Skliarevskii} \& E.~I. {Vorobyov}(2023){Skliarevskii} \& {Vorobyov}}]{Skliarevskii2023}
{Skliarevskii}, A.~M., \& {Vorobyov}, E.~I. 2023, \bibinfo{title}{{Luminosity Outbursts in Interacting Protoplanetary Systems},} Astronomy Reports, 67, 1401, \dodoi{10.1134/S1063772923120107}

\bibitem[{J.~M. {Stone} \& M.~L. {Norman}(1992){Stone} \& {Norman}}]{SN1992}
{Stone}, J.~M., \& {Norman}, M.~L. 1992, \bibinfo{title}{{ZEUS-2D: A Radiation Magnetohydrodynamics Code for Astrophysical Flows in Two Space Dimensions. I. The Hydrodynamic Algorithms and Tests},} \apjs, 80, 753, \dodoi{10.1086/191680}

\bibitem[{M. {Takami} {et~al.}(2018){Takami}, {Fu}, {Liu}, {Karr}, {Hashimoto}, {Kudo}, {Vorobyov}, {K{\'o}sp{\'a}l}, {Scicluna}, {Dong}, {Tamura}, {Pyo}, {Fukagawa}, {Tsuribe}, {Dunham}, {Henning}, \& {de Leon}}]{Takami:2018aa}
{Takami}, M., {Fu}, G., {Liu}, H.~B., {et~al.} 2018, \bibinfo{title}{{Near-infrared High-resolution Imaging Polarimetry of FU Ori-type Objects: Toward a Unified Scheme for Low-mass Protostellar Evolution},} \apj, 864, 20

\bibitem[{R. {Teague} {et~al.}(2018){Teague}, {Bae}, {Bergin}, {Birnstiel}, \& {Foreman-Mackey}}]{Teague:2018aa}
{Teague}, R., {Bae}, J., {Bergin}, E.~A., {Birnstiel}, T., \& {Foreman-Mackey}, D. 2018, \bibinfo{title}{{A Kinematical Detection of Two Embedded Jupiter-mass Planets in HD 163296},} \apjl, 860, L12

\bibitem[{R. {Teague} {et~al.}(2025){Teague}, {Benisty}, {Facchini}, {Fukagawa}, {Pinte}, {Andrews}, {Bae}, {Barraza-Alfaro}, {Cataldi}, {Cuello}, {Curone}, {Czekala}, {Fasano}, {Flock}, {Galloway-Sprietsma}, {Garg}, {Hall}, {Hammond}, {Hilder}, {Huang}, {Ilee}, {Izquierdo}, {Kanagawa}, {Lesur}, {Lodato}, {Longarini}, {Loomis}, {Masset}, {Menard}, {Orihara}, {Price}, {Rosotti}, {Stadler}, {Testi}, {Yen}, {Wafflard-Fernandez}, {Wilner}, {Winter}, {W{\"o}lfer}, {Yoshida}, \& {Zawadzki}}]{Teague:2025aa}
{Teague}, R., {Benisty}, M., {Facchini}, S., {et~al.} 2025, \bibinfo{title}{{exoALMA. I. Science Goals, Project Design, and Data Products},} \apjl, 984, L6

\bibitem[{E.~I. {Vorobyov} {et~al.}(2018){Vorobyov}, {Akimkin}, {Stoyanovskaya}, {Pavlyuchenkov}, \& {Liu}}]{2018VorobyovAkimkin}
{Vorobyov}, E.~I., {Akimkin}, V., {Stoyanovskaya}, O., {Pavlyuchenkov}, Y., \& {Liu}, H.~B. 2018, \bibinfo{title}{{Early evolution of viscous and self-gravitating circumstellar disks with a dust component},} \aap, 614, A98, \dodoi{10.1051/0004-6361/201731690}

\bibitem[{E.~I. {Vorobyov} \& S. {Basu}(2005){Vorobyov} \& {Basu}}]{VorobyovBasu2005}
{Vorobyov}, E.~I., \& {Basu}, S. 2005, \bibinfo{title}{{The Origin of Episodic Accretion Bursts in the Early Stages of Star Formation},} \apjl, 633, L137, \dodoi{10.1086/498303}

\bibitem[{E.~I. {Vorobyov} \& S. {Basu}(2009){Vorobyov} \& {Basu}}]{VorobyovBasu2009}
{Vorobyov}, E.~I., \& {Basu}, S. 2009, \bibinfo{title}{{Secular evolution of viscous and self-gravitating circumstellar discs},} \mnras, 393, 822, \dodoi{10.1111/j.1365-2966.2008.14376.x}

\bibitem[{E.~I. {Vorobyov} \& S. {Basu}(2015){Vorobyov} \& {Basu}}]{VorobyovBasu2015}
{Vorobyov}, E.~I., \& {Basu}, S. 2015, \bibinfo{title}{{Variable Protostellar Accretion with Episodic Bursts},} \apj, 805, 115, \dodoi{10.1088/0004-637X/805/2/115}

\bibitem[{E.~I. {Vorobyov} \& V.~G. {Elbakyan}(2018){Vorobyov} \& {Elbakyan}}]{2018VorobyovElbakyan}
{Vorobyov}, E.~I., \& {Elbakyan}, V.~G. 2018, \bibinfo{title}{{Gravitational fragmentation and formation of giant protoplanets on orbits of tens of au},} \aap, 618, A7, \dodoi{10.1051/0004-6361/201833226}

\bibitem[{E.~I. {Vorobyov} {et~al.}(2021){Vorobyov}, {Elbakyan}, {Liu}, \& {Takami}}]{Vorobyov2021}
{Vorobyov}, E.~I., {Elbakyan}, V.~G., {Liu}, H.~B., \& {Takami}, M. 2021, \bibinfo{title}{{Distinguishing between different mechanisms of FU-Orionis-type luminosity outbursts},} \aap, 647, A44, \dodoi{10.1051/0004-6361/202039391}

\bibitem[{P. {Weber} {et~al.}(2023){Weber}, {P{\'e}rez}, {Zurlo}, {Miley}, {Hales}, {Cieza}, {Principe}, {C{\'a}rcamo}, {Garufi}, {K{\'o}sp{\'a}l}, {Takami}, {Kastner}, {Zhu}, \& {Williams}}]{Weber:2023ab}
{Weber}, P., {P{\'e}rez}, S., {Zurlo}, A., {et~al.} 2023, \bibinfo{title}{{Spirals and Clumps in V960 Mon: Signs of Planet Formation via Gravitational Instability around an FU Ori Star?},} \apjl, 952, L17

\bibitem[{P. {Weber} {et~al.}(2025){Weber}, {Ulloa}, {P{\'e}rez}, {Miley}, {Cieza}, {Nayakshin}, {Zurlo}, {Liu}, {Cruz-S{\'a}enz de Miera}, {Hales}, {Garufi}, {Stamatellos}, {K{\'o}sp{\'a}l}, \& {Guzm{\'a}n}}]{Weber2025}
{Weber}, P., {Ulloa}, S., {P{\'e}rez}, S., {et~al.} 2025, \bibinfo{title}{{A Multiwavelength Study of the Dynamic Environment Surrounding the FUor V960 Mon},} \apj, 985, 226, \dodoi{10.3847/1538-4357/adc9a2}

\bibitem[{H.~W. {Yorke} \& P. {Bodenheimer}(2008){Yorke} \& {Bodenheimer}}]{2008YorkeBodenheimer}
{Yorke}, H.~W., \& {Bodenheimer}, P. 2008, in Astronomical Society of the Pacific Conference Series, Vol. 387, Massive Star Formation: Observations Confront Theory, ed. H.~{Beuther}, H.~{Linz}, \& T.~{Henning}, 189

\bibitem[{Z. {Zhu} {et~al.}(2010){Zhu}, {Hartmann}, {Gammie}, {Book}, {Simon}, \& {Engelhard}}]{Zhu2010}
{Zhu}, Z., {Hartmann}, L., {Gammie}, C.~F., {et~al.} 2010, \bibinfo{title}{{Long-term Evolution of Protostellar and Protoplanetary Disks. I. Outbursts},} \apj, 713, 1134, \dodoi{10.1088/0004-637X/713/2/1134}

\end{thebibliography}
\bibliographystyle{aasjournalv7}

\end{document}